\documentclass[12pt]{iopart}

\usepackage{epsfig}
\usepackage{graphicx}
\usepackage{dcolumn}
\usepackage{bm}
\usepackage{bbm}
\usepackage{wasysym}
\usepackage{marvosym}
\usepackage{times}
\usepackage{amssymb}
\usepackage{subfigure}
\newcommand{\abs}[1]{{\left|#1\right|}}
\newcommand{\sgn}[1]{{\mathrm {\,sgn}(#1)}}

\def\Xint#1{\mathchoice
  {\XXint\displaystyle\textstyle{#1}}%
  {\XXint\textstyle\scriptstyle{#1}}%
  {\XXint\scriptstyle\scriptscriptstyle{#1}}%
  {\XXint\scriptscriptstyle\scriptscriptstyle{#1}}%
  \!\int}
\def\XXint#1#2#3{{\setbox0=\hbox{$#1{#2#3}{\int}$}
  \vcenter{\hbox{$#2#3$}}\kern-.5\wd0}}

\def\dashint{\Xint-}

\begin{document}

\title{Local density of states and scanning tunneling currents in graphene }

\author{N.~M.~R. Peres$^1$, Ling Yang$^2$ and Shan-Wen Tsai$^2$}

\address{$^1$
  Departament of
Physics and Center of Physics, University of Minho, P-4710-057, Braga, Portugal
\ead{peres@fisica.uminho.pt}}
\address{
$^2$
Department of Physics and Astronomy, University of California, 
Riverside, CA 92521, USA
\ead{ling.yang@email.ucr.edu}
\ead{swtsai@physics.ucr.edu}
}

\pacs{73.20.Hb, 73.23.-b, 81.05.Uw}
\begin{abstract}
 We present exact analytical calculations of scanning tunneling currents
in locally disordered graphene using a multimode description of the microscope
tip. Analytical expressions for the local density of states (LDOS) are given
for energies beyond the Dirac cone approximation. We show that the
LDOS at the $A$ and $B$ sublattices of graphene are out of phase by $\pi$ implying
that the averaged LDOS, as one moves away from the impurity, shows no trace
of the $2q_F$ (with $q_F$ the Fermi momentum) Friedel modulation. This means
that a STM experiment lacking atomic resolution at the sublattice level will
not be able of detecting the presence of the Friedel oscillations [this seems to be the
case in the experiments reported in Phys. Rev. Lett. {\bf 101}, 206802 (2008)].
The momentum maps of the LDOS for different types of impurities are given. In the
case of the vacancy, $2q_F$ features are seen in these maps. In all momentum space
maps, $K$ and $K+K^\prime$ features are seen. The $K+K^\prime$ features are 
different from what is seen around zero momentum.
An interpretation for 
these features
is given.
The calculations reported here
are valid for chemical substitution impurities, such as boron and nitrogen atoms, as well
as for vacancies. It is shown
that the density of states close to the impurity is very sensitive to type of disorder:
diagonal, non-diagonal, or vacancies.
In the case of weakly coupled (to the carbon atoms) impurities,  
the local density of states presents strong resonances at finite energies, 
which leads to steps in the scanning tunneling currents and to suppression of the Fano factor.
\end{abstract}

\NJP


\section{Introduction}

Graphene\cite{novo1,pnas} consists of a monolayer of covalently bonded carbon atoms 
forming a two-dimensional honeycomb lattice\cite{rmp,EPN}. Low-energy electronic excitations in graphene are well described as massless Dirac fermions with an additional pseudospin degree of freedom. Because of the Dirac spectrum, impurities can have a strong effect on the local electronic structure of graphene when the Fermi energy is near the Dirac point\cite{Peres2006,pereira,loktev,cheianov,bena,Peres2007,charlier}. Impurities in graphene can be in the substrate, in the form of adatoms, or as imperfections in the lattice itself. In one hand, there has been very significant progress in decreasing the amount of disorder introduced in graphene, for example by fabrication of suspended samples\cite{Meyer}. On the other hand, impurity effects have been explored to modify and tailor the electronic, thermal and chemical properties of graphene. Examples of the later include experiments with graphane\cite{novoselov}, chemical substitution of some of graphene's carbon atoms by boro!
 n an!
 d nitrogen atoms\cite{peng,Panchakarla}, and doping of graphene with metals on top\cite{Chen,Pi}. 

Since graphene is an atomically thin membrane, it can be easily accessed with Scanning Tunneling Microscopy (STM) measurements. Impurity effects can be studied with atomic resolution and the local spectrum can be obtained by STM spectroscopy. In addition, atomic manipulation can also be performed with STM. There has been several STM studies of graphene grown epitaxially on SiC \cite{stm_rutter,stm_mallet,stm_brar}, mechanically exfoliated graphene on SiO$_2$ \cite{stm_ishigami,substrate,stm_geringer,stm_zhang,stm_deshpande}, and graphene flakes on graphite\cite{stm_andrei}. In fact, STM experiments have proved instrumental in mapping the
topography of corrugated graphene and
determining the
existence of charge puddles \cite{cromie}. Additionally, 
STM experiments are also able to probe the chiral nature of the electrons in graphene
 when they scatter from impurities  \cite{Bose}. This experimental work showed that
intravalley backscattering is virtually absent in graphene. In particular, the STM experiment showed
the lack  of the $2q_F$ ($q_F$ is the Fermi momentum) Friedel 
modulation on the local density of states
of graphene. As we show explicitly below, this lack of modulation
can be traced to the fact that the local density of states at the $A$ and $B$ sublattices
are out of phase by $\pi$, an aspect already noted in passing previously  \cite{bena2,barnea}. 
Therefore, the local density of states, when averaged over the
unit cell, shows no trace of the $2q_F$ oscillation. Additionally, as we show below,
at distances $d$ close to the impurity, $d\ll 1/q_F$, there is a strong departure
from the $1/r^2$ spatial dependence \cite{balatsky,cheianov} 
of the local density of states.
Moreover the form
of the density of states close to the impurity is very sensitive to type of disorder:
diagonal, non-diagonal, or vacancies.

In this work, we present calculations of STM currents in locally disordered graphene. We focus on the case of chemical substitution (by boron or nitrogen atoms, for example) and use a multimode description for the STM tip. We obtain exact analytical expressions for the local density of states, and also present results for energies beyond the Dirac cone approximation. 
We model the substitutional impurity by both an on-site impurity potential and local hopping disorder. We find that inclusion of the hopping disorder term leads to additional higher harmonics oscillations in real space for the density of states for the sub-lattice that does not contain the impurity. The main oscillations in the two sub-lattices are out-of-phase away from the impurity.
For the regime in which the electronic hopping between the impurity and the nearest neighbor carbon atoms is decreased in relation to the hopping between carbon atoms in the clean system, the local density of states presents strong resonances. A vacancy is a extreme case of this regime. These resonances lead to the appearance of steps in the STM current, a signature that should be observable experimentally. These resonances also leads to open channels for tunneling between the STM tip and graphene, and lead to a decrease in the Fano factor. 

Another issue addressed in this paper relates to the effect of the tip
on the measured STM currents. In  the usual analysis, the electrons
in the tip are represented by jellium model with constant density of states. In this type
of model neither the real part of the self-energy due to the
tip-system coupling nor the variation of the density of states with energy is included
(wide band limit).
The tip, however, is not an infinite metal. In fact it has a structure where
the number of atoms in the atomic planes reduces as we approach the tip.
In a previous publication \cite{stmperes} we have modeled the tip as a one dimensional
model (in that work we have also considered the simplification of zero on-site energy
at the impurity), which corresponds essentially to the case of a constant
density of state too a good approximation. In that case we found the
STM current to be symmetric around zero energy. When we generalize to the
case of  a multimode tip this symmetry is lost, as we show in this work.
Comparing the results of Ref. \cite{stmperes} with those given here
it is possible to disentangle the effects due to graphene and to
the impurities from those due to the tip. This work shows that some care
has to be taken when interpreting the STM currents directly.

The present manuscript is organized in the following way: In Section 2, the Green's function formalism is presented, with analytical results for the Green's function for graphene with a substitutional impurity, and for the STM tip modeled by a multimode system. Local density of states results are presented in Section 3, and results for the STM current are presented in Section 4. Section 5 contains final discussions and conclusions.

\section{Graphene and STM tip Green's functions}
\subsection{Graphene}
\label{Graphene}

\begin{figure}[htf]
\begin{center}
\includegraphics*[width=7cm]{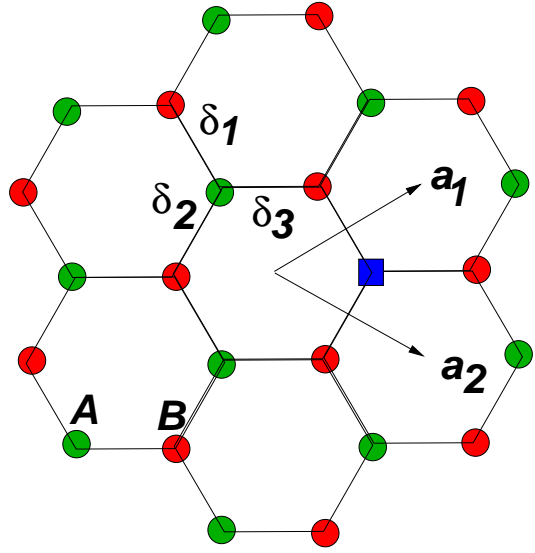}
\end{center}
\caption{(color on line)
Honeycomb lattice of graphene, with an substituting impurity at the
$A$ sublattice (the square). The unit cell vector $\bm a_1$ and
$\bm a_2$ as well as the next nearest neighbors vectors
$\bm\delta_i$ ($i=$1,2,3) are also represented.
 \label{Fig_lattice}}
\end{figure}

The honeycomb lattice has two carbon atoms per unit cell, one from sublattice A and one from sublattice B, as depicted in Fig.~\ref{Fig_lattice}. 
The unit cell vectors are $\bm a_1$ and $\bm a_2$, with magnitudes  
$\vert \bm a_1 \vert=\vert \bm a_2 \vert=a$, where $a = \sqrt{3} a_0 \simeq 2.461$ \AA, and $a_0$
is the carbon-carbon distance. Any lattice vector $\bm r$
can be represented in this basis as
$\bm r= n\bm a_1 + m\bm a_2$, with $n,m$ integers. 
In Cartesian coordinates,
$\bm a_1 = a_0 (3,\sqrt 3,0)/2$ and 
$\bm a_2 = a_0 (3,-\sqrt 3,0)/2$, and 
the reciprocal lattice vectors are given by:
$\bm b_1= 2\pi (1,\sqrt 3,0)/(3 a_0)$ and 
$\bm b_2= 2\pi (1,-\sqrt 3,0)/(3 a_0)$.
The  vectors connecting any $A$ atom to its nearest
neighbors are: 
$\bm \delta_1 = (\bm a_1-2\bm a_2)/3$, 
$\bm \delta_2 = (\bm a_2-2\bm a_1)/3$, and 
$\bm \delta_3 = (\bm a_1+\bm a_2)/3$.

We consider here the case where a substituting atom replaces a carbon atom
in the $A$ sublattice, say. When this happens two effects take place:
(i) the on-site energy $\epsilon_i$ at the impurity site is different from that at the
carbon atoms; (ii) the hopping from and to the impurity atom, $t_i$, changes relatively
to that of pristine graphene. In the latter case, we model the change
in the hopping by introducing an additional non-diagonal term
to the Hamiltonian, such that $t_i=-t+t_0$ (see below). 
Using these 
definitions the Hamiltonian can be written as: $H=H_0+V_t+V_i$, where 

\begin{eqnarray}
H_0 &=& -t \sum_{\bm r} [b^\dag(\bm r) a(\bm r)
+ b^\dag (\bm r - \bm a_2) a(\bm r)
+b^\dag(\bm r - \bm a_1) a(\bm r) + {\rm h.c.}],
\end{eqnarray}
is the kinetic energy operator and $a^\dag$, $a$ ($b^\dag$, $b$) are fermion creation and annihilation operators in the $A$ ($B$) sites. The spin index is omitted for
simplicity. We consider an isolated impurity located at ${\bm r} = (0, 0, 0)$, on sublattice A, so that its contribution to the Hamiltonian has two terms:
\begin{equation}
\label{Vi}
V_t = t_0[b^\dag(0) a(0) + b^\dag(-\bm a_2)a(0) + 
b^\dag(-\bm a_1) a(0) + {\rm h.c.}]
\end{equation}
 and 
\begin{equation}
\label{V0}
V_i=\varepsilon_i a^{\dagger}(0)a(0) \ ,
\end{equation}
for hopping and potential disorder, respectively. 
In the case of zero chemical potential, when the Fermi
 level crosses the Dirac point, the system is most susceptible to the presence of impurities. 
In the particular case $t_0=t$, hopping to the impurity site is completely suppressed, and the scattering term
$V_t$ represents a vacancy. 
It is well known\cite{barbary} that the formation 
of a vacancy will lead to
some local distortion of the carbon-carbon bonds. This effect is not
incorporated in our Hamiltonian, which in that case would not be
exactly solvable. The substitution of carbon atoms by boron or
nitrogen has the main consequence  of changing  the local hopping
and the onsite energy,
both effects included in our description. The particular choice
for boron or nitrogen is due to size restrictions imposed
by the unit cell of graphene. We note here that replacement of carbon
atoms by boron and nitrogen was already experimentally achieved \cite{Panchakarla}.

We first calculate the single particle Green's functions for the system comprised of graphene and a single impurity, described by the Hamiltonian $H$ above. The single particle Green's functions carry sub-lattice indices, and are defined as: 
\begin{eqnarray}
G_{aa}(\bm k,\bm q,\tau) &=&
   -\left< T \left[a_{\bm k}(\tau) \: a^\dag_{\bm q}(0)\right] \right>\,,\\
G_{bb}(\bm k,\bm q,\tau) &=&
   -\left< T \left[b_{\bm k}(\tau) \: b^\dag_{\bm q}(0)\right] \right>\,,\\
G_{ab}(\bm k,\bm q,\tau) &=&
   -\left< T \left[a_{\bm k}(\tau) \: b^\dag_{\bm q}(0)\right] \right>\,,\\
G_{ba}(\bm k,\bm q,\tau) &=&
   -\left< T \left[b_{\bm k}(\tau) \: a^\dag_{\bm q}(0)\right] \right>\,.
\end{eqnarray}
The equations of motion for the Green's functions are given by:
\begin{eqnarray}
  i\omega_n G_{aa} (\omega_n, \bm k, \bm p )   &=&   \delta_{\bm k,\bm p}   +    \sum_{\bm q}   \left[  \lambda_{\bm k, \bm q} G_{ba} (\omega_n, \bm q,  \bm p )  +   \frac{\epsilon_i}{N_c} G_{aa} (\omega_n, \bm q, \bm p ) \right] \\
  i\omega_n G_{ba} (\omega_n, \bm k, \bm p )   &=&    \sum_{\bm q} 
\lambda^{\ast}_{\bm q, \bm k} G_{aa}(\omega_n,\bm q,\bm p )\\
  i\omega_n G_{ab} (\omega_n, \bm k, \bm p )   &=&    \sum_{\bm q}   \left[ \lambda_{\bm k,\bm q} G_{bb}(\omega_n,\bm q,\bm p ) +   \frac{\epsilon_i}{N_c} G_{ab} (\omega_n, \bm q, \bm p ) \right] \\
  i\omega_n G_{bb} (\omega_n, \bm k, \bm p )   &=&    \delta_{\bm
  k,\bm p}  +  \sum_{\bm q} \lambda^{\ast}_{\bm q,\bm k} G_{ab}(\omega_n,\bm q,\bm
p ) \, ,
\end{eqnarray}
where 
\begin{equation}
\lambda_{\bm k,\bm p} = -t\phi_{\bm p} (\delta_{\bm k, \bm p} - t_0/N_c t)\,,
\end{equation}
\begin{equation}
\phi_{\bm p} =1+e^{-i\bm p\cdot \bm a_1}+e^{-i\bm p\cdot \bm a_2}\,,
\end{equation}
 and $N_c$
is the total number of unit cells in the lattice, and $\omega_n$ are fermionic Matsubara frequencies. 
Note that $\lambda_{\bm p\bm q}\ne \lambda_{\bm q\bm p}$. This is a consequence of the impurity hopping term $V_t$, which breaks sub-lattice symmetry. The impurity potential term $V_i$ also breaks sub-lattice symmetry, and therefore $\epsilon_i$ appears in an asymmetric way in the equations above. The set of equations of motions can be solved exactly. The presence of the
scattering term $V_t$ leads to the appearance of the phases $\phi_{\bm k}$ and a more complex form for the $T$-matrix than usual. 
The exact solution for the Green's functions can be written, after
a lengthy calculation, in the form\cite{Peres2007}:
\begin{eqnarray}  
G_{aa}(\bm k,\bm p)  &=&  \delta_{\bm k,\bm p} \ G^0_{\bm k} + g +
   h \left[ G^0_{\bm k} + G^0_{\bm p} \right] 
+ G^0_{\bm k} \ T \ G^0_{\bm p} \;,
\label{gaa}\\
  G_{bb}(\bm k,\bm p)  &=&  \delta_{\bm k,\bm p} \ G^0_{\bm k} +
    \frac{t \phi^*_{\bm k}}{i\omega_n} \ G^0_{\bm k} \ T \ 
    G^0_{\bm p} \ \frac{t \phi_{\bm p}}{i\omega_n} \;.
\label{gbb}
\end{eqnarray}
where all the terms ($G$, $G^0$, $g$, $h$ and $T$) also depend on $\omega_n$ (omitted here for brevity).  The terms $g$, $h$, and $T$ correspond to sums over infinite series of Feynman diagrams for impurity scattering, and are given by: 
\begin{eqnarray} 
g(\omega_n)&=& t_0^2 \bar{G}^0(\omega_n)/[N_c D(\omega_n)] , \\
h(\omega_n) &=&  t_0(t-t_0)/[N_c D(\omega_n)]  ,
\label{multiscattering}
\end{eqnarray}
and 
\begin{eqnarray}
T(\omega_n) =
   -\frac{i\omega_n t_0 (2t-t_0) - \epsilon_i t^2}{N_c D(\omega_n)} 
\end{eqnarray}
where the denominator $D(\omega_n)$ is defined as
\begin{eqnarray}
  D(\omega_n)  =  (t - t_0)^2  + 
  \left[ i\omega_n t_0 (2t - t_0)  -  \epsilon_i t^2 \right] 
  \bar{G}^0(\omega_n)
\label{Domega}
\end{eqnarray}
and 
\begin{equation}
\bar{G}^0(\omega_n) =\frac 1 {N_c}\sum_{\bm k} G^0(\omega_n, \bm k)
\label{gbar}
\end{equation}
 with the diagonal component of the Green's function for the clean system given by ($\hbar=1$)
\begin{equation}
G^0_{\bm k}=  G^0(\omega_n, \bm k) =  
\frac {i\omega_n}{(i\omega_n)^2-t^2|\phi_{\bm k}|^2} \ ,
\label{green}
\end{equation}
which is translationally invariant. The expressions for the Green's functions,
Eqs. (\ref{gaa})-(\ref{Domega}), are exact analytic solutions for graphene with one isolated substitutional impurity, including contributions from both 
the on-site energy $\epsilon_i$ and the hopping parameter $t_0$. Inclusion of the off-diagonal disorder $t_0$ leads to additional terms, and additional $\omega$-dependence of the graphene Green's function. Since single particle properties, such as local electronic spectra, can be obtained directly from the Green's functions, this $\omega$-dependence has direct experimental consequences, such as for STM spectroscopy measurements.  
These results for the graphene Green's functions have been obtained in Ref. \cite{Peres2007}, where local density of states maps, electronic spectra and Friedel oscillations have been calculated for the cases of boron and nitrogen substitution. We include here a brief derivation of the analytical expressions for the Green's functions, Eqs. (\ref{gaa})-(\ref{green}), for completeness. 
Upon closer inspection, the physical meaning of the extra terms in $G_{aa}$ become clear.
The significance of the term $g(\omega_n)$ in (\ref{gaa}) which only appears in $G_{aa}$, 
is more easily interpreted if we do a double Fourier transform to real space. 
This term corresponds to the return amplitude 
to the impurity site for an electron starting at the impurity site.
The factor 
$1/D(\omega_n)$ contains a sum over an infinite series of intermediate 
scattering events, but the overall process is bounded and the $t_0^2$ factor 
denotes hopping from the impurity to the nearest neighbor $B$-sites and back to the impurity site.  
Likewise, an interpretation can be given to the other term which only appears in
$G_{aa}$, namely, $h(\omega_n) G^{0}_{\bm k}(\omega_n)$. A double Fourier transform shows that this term contributes to $G_{aa}({\bm r},0)$ and describes the amplitude of propagation between the 
impurity site and another $A$ site, again with an infinite series of intermediate scatterings. Similarly, the $h(\omega_n) G^{0}_{\bm p}(\omega_n)$ term contributes to $G_{aa}(0,{\bm r})$. No such terms can, of course, appear in $G_{bb}$
when the inpurity is at a $A$ site. And no such term can be present when there is only the impurity potential term $\epsilon_i$. The $G^0_{\bm k}(\omega_n) T(\omega_n) G^0_{\bm p}(\omega_n)$ term, which appears in both $G_{aa}$ and $G_{bb}$, is the usual term also present in simple on-site impurity potential problems, but in this case the $T$-matrix contains contributions from both $t_0$ and $\epsilon_i$.

\subsection{STM Tip}
\label{tip}

\begin{figure}[ht]
\begin{center}
\includegraphics*[angle=0,width=8cm]{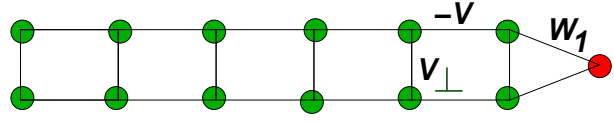}
\caption{(colour online)
Representation of the STM tip.
 \label{Fig:tip}}
\end{center}
\end{figure}

Let us consider a model for the STM tip represented by a multimode system.
The bulk of the tip is modeled by a square lattice with two atoms in the transverse
direction. The end of the tip is represented by a single atom.
This choice renders the system multimode, with two
transverse modes. It is as simple to include a truly three dimensional tip, but the
current will not be much affected by it. The schematic atomic structure of the
tip is represented in Fig. \ref{Fig:tip}.

\begin{figure}[ht]
\begin{center}
\includegraphics*[angle=0,width=8cm]{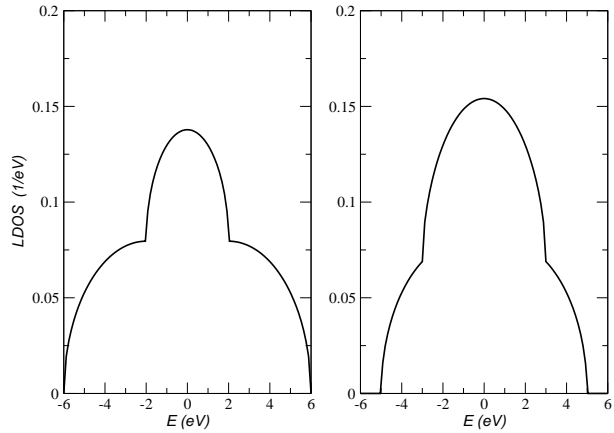}
\caption{Local density of states, $\rho_b(E)$, at the atoms of the tip
given by $n=-1,m=1,2$. Left: $V=2$ eV, $V_\perp=2$ eV.
Right: $V=2$ eV, $V_\perp=1$ eV. The multi-mode nature of the tip
is clearly seen.
 \label{Fig:ldos_tip}}
\end{center}
\end{figure}

The Hamiltonian for the tip can be written as $H_t=H_b+H_0$, 
where $H_b$ represents the bulk of the tip and $H_0$ the tip's last atom.
These two parts of the Hamiltonian are defined as 
\begin{eqnarray}
 H_b=&&-V\sum_{n=-\infty}^{-1}\sum_{m=1,2}[c^\dag(n,m)c(n-1,m)+c(n-1,m)c^\dag(n,m)]\\
&&-V_\perp\sum_{n=-\infty}^{-1}[c^\dag(n,1)c(n,2)+c(n,2)c^\dag(n,1)]\,,
\end{eqnarray}
and 
\begin{equation}
 H_0=\epsilon_0c^\dag(0)c(0) -W_1 \sum_{m=1,2}[c^\dag(0)c(-1,m)+c(-1,m)c^\dag(0)] \, ,
\end{equation}
where $c^{\dagger}$ ($c$) are creation (annihilation) operators for fermions in the tip.
Let us now consider the case of the bulk part of the Hamiltonian's tip, $H_b$.
In the case of a square lattice the wave function is separable and can be written
as
 $\vert\psi_{l,t}\rangle = \vert\phi_l\rangle \vert\phi_t\rangle$,
with the longitudinal part of the wave function  $\vert\phi_l\rangle$ given by
\begin{equation}
\vert \phi_l\rangle=\lim_{N\rightarrow\infty}\sum_{n=-1}^{-N}\sqrt{\frac{2}{N+1}}\sin(n\theta_l)\vert n\rangle  \  ,
\label{phil}
\end{equation}
and the transverse part $\vert\phi_t\rangle$ given by 
\begin{equation}
 \vert\phi_t\rangle=\sum_{m=1,2}\sqrt{\frac{2}{3}}\sin(m\alpha_t)\vert m\rangle.
\label{phit}
\end{equation}
In Eqs. (\ref{phil}) and (\ref{phit}), the states 
$\vert n,m \rangle=\vert n \rangle\vert m \rangle$ are position states and the
numbers $\theta_l$ and $\alpha_t$ are given by
\begin{equation}
 \theta_l = \frac{\pi l}{N+1}\,,\hspace{0.5cm} l=1,2,\ldots,N\,,
\end{equation}
and
\begin{equation}
 \alpha_t = \frac{\pi t}{3}\,,\hspace{0.5cm} t=1,2\,.
\end{equation}
The resolvent for the Hamiltonian $H_b$ is defined as
 $\hat G_b^+=(E+i0^+-H_b)^{-1}$,
where the $+$
superscript denotes the retarded function. In the eigenstate basis, it has the form
\begin{equation}
 \hat G_b^+=\sum_{l,t}\frac{\vert \psi_{l,t}\rangle\langle \psi_{l,t} \vert}{E-E_{l,t}}\,,
\end{equation}
where $E_{l,t}$ are the eigenvalues of $H_b$, 
with $H_b\vert \psi_{l,t}\rangle=E_{l,t}\vert \psi_{l,t}\rangle$, given
by
\begin{equation}
 E_{l,t}=-2V\cos\theta_l-2V_\perp\cos\alpha_t\,,
\end{equation}
For the calcution
of the STM current we will need the surface Green's functions defined as
\begin{eqnarray}
\label{Gdiag}
G_{diag}(E) =\langle m,-1 \vert G_b^+ \vert -1,m\rangle\,,\\
\label{Goffd}
G_{offd}(E) = \langle 1,-1 \vert G_b^+ \vert -1,2\rangle\,.
\end{eqnarray}
The calculation of (\ref{Gdiag}) and (\ref{Goffd}) requires the
evaluation of the integral
\begin{equation}
 I=\frac{1}{2\pi}\int_0^{2\pi}\frac{d\,\theta\sin^2\theta}{E+2V\cos\theta+jV_\perp}
,\hspace{0.5cm}j=\pm1\,,
\end{equation}
which is easily done by contour integration methods \cite{peresGreen}.
The final results are 
\begin{eqnarray}
\label{Gdiagf}
G_{diag}(E)&=&\sum_{j=\pm 1} \frac{\beta_j}{2V}-\sgn{\beta_j}\frac{1}{2V}\sqrt{\beta^2_j-1}\,,\\
\label{Goffdf}
 G_{offd}(E)&=&\sum_{j=\pm 1} \frac{j\beta_j}{2V}-\sgn{\beta_j}j\frac{1}{2V}\sqrt{\beta^2_j-1}\,,
\end{eqnarray}
for $\beta_j^2>1$, with $\beta_j=(E+jV_\perp)/(2V)$. In the case $\beta^2_j<1$, the Green's functions are obtained
from (\ref{Gdiagf}) and (\ref{Goffdf}) by removing the factor $\sgn{\beta_j}$, and choosing the
positive sign for the square root of the negative argument:
\begin{eqnarray}
\label{GdiagfI}
G_{diag}(E)&=&\sum_{j=\pm 1} \frac{\beta_j}{2V}-i\frac{1}{2V}\sqrt{1-\beta^2_j}\,,\\
\label{GoffdfI}
 G_{offd}(E)&=&\sum_{j=\pm 1} \frac{j\beta_j}{2V}-ji\frac{1}{2V}\sqrt{1-\beta^2_j}\,,
\end{eqnarray}
Using Eq. (\ref{Gdiagf}), the local density of states at the sites $n=-1,m=1,2$
given as usual by
 $\rho_b(E)=-\frac 1 {\pi}\Im G_{diag}(E)$,
is depicted in Fig. \ref{Fig:ldos_tip}. The multimode nature
of the tip is clearly seen in the form of the density of states. 

\section{Graphene's local density of states}

The properties of the STM current depend on the local density of states below the tip of the microscope. In this section we compute the Green's function of graphene in real space, from which the local density of states can be obtained.
In Sec. \ref{Graphene}, the position vector $\bm r$ denotes the position of the unit cell, which contains two atoms. The local density of states (per spin) at the sub-lattice $A$ and sub-lattice $B$
atoms of the unit cell
localized in the position $\bm r$ is defined as 
\begin{equation}
\rho_x(\bm r,\omega) = -\frac 1 {\pi N_c} 
{\rm Im} G_{xx}(\bm r,\bm r,\omega)\,,
\label{rho_x}
\end{equation}
where $x=a,b$, and $G_{xx}(\bm r,\bm r,\omega)$ is obtained  from
\begin{eqnarray}
G_{xx}(\bm r,\bm r,\omega_n) &=& 
\sum_{\bm k,\bm p}e^{i(\bm k-\bm p)\cdot \bm r}
 G_{xx}(\bm k,\bm p, \omega_n)\,,
\label{realspaceG}
\end{eqnarray}
after the usual analytical 
continuation $\omega_n\rightarrow \omega + i0^+$ of the Matsubara Green's function.
In the unit cell $\bm r=0$, which contains the impurity in its $A$ site, we have simple expressions for the Green's functions for the $A$ and $B$ sites. These read
\begin{eqnarray}
 G_{aa}(\omega_n)=\bar G^0(\omega_n)+g(\omega_n)+2h(\omega_n)\bar G^0(\omega_n)
+[\bar G^0(\omega_n)]^2T(\omega_n)\,,\\
G_{bb}(\omega_n)=\bar G^0(\omega_n)+\frac{t^2}{9(i\omega_n)^2}[\tilde G^0(\omega_n)]^2T(\omega_n)\,,
\end{eqnarray}
where $\tilde G^0(\omega_n)$ is given by
\begin{equation}
 \tilde G^0(\omega_n)=-i\omega_nt^{-2}+(i\omega_n)^2t^{-2}\bar G^0(\omega_n) \ ,
\end{equation}
and $\bar G^0(\omega_n)$ is defined in Eq. (\ref{gbar}).
As is clear from the above equations, the central quantity that needs to be calculated is the
integrated Green's function $\bar G^0(\omega_n)$. Since we want to compute the
density of states for energies beyond the Dirac cone approximation we need
to include in the density of states powers of the energy beyond the usual
linear term. In a previous work\cite{condgeim08},  
we have derived an expansion for the density of states (per unit cell, per spin)
valid for energies up to
$\sim 2.5$ eV, reading ($E=\hbar\omega$)
\begin{equation}
\rho(E)\simeq \frac {2E}{\sqrt 3 \pi t^2}+\frac {2E^3}{3\sqrt 3 \pi t^4}
+\frac {10E^5}{27\sqrt 3 \pi t^6}\,.
\label{expand_rho}
\end{equation}
Using this expression for the density of states, a close form for the
retarded function $\bar G^0(\omega_n\rightarrow \omega +i0^+)$ can be
derived.
The imaginary part of $\bar G^0(\omega)$ reads
\begin{equation}
 \Im \bar G^0(\omega) = -\frac{\pi}2\rho(\hbar\omega)\,,
\label{imGbar}
\end{equation}
and the real part has the form
\begin{equation}
 \Re \bar
G^0(\omega)=P_1(\hbar\omega)+P_2(\hbar\omega)\ln\frac{(\hbar\omega)^2}{D_c^2-(\hbar\omega)^2}\,,
\label{reGbar}
\end{equation}
where 
$P_1(x)$ and $P_2(x)$ are polynomial functions given by
\begin{eqnarray}
 P_1(x)&=&-\frac {x}{3t^2}-\frac{5}{27t^4}\left(
\frac x 2 D^2_c + x^3
\right)\,,\\
P_2(x)&=&\frac {x}{D_c^2} + \frac{x^3}{3t^2D^2_c} + \frac{5}{27D^2_ct^4}x^5\,.
\end{eqnarray}
The energy $D_c$ is a cut-off energy chosen as $D^2_c=\sqrt 3\pi t^2$.
It is possible to derive simple analytical expressions for the local density of states
at the unit cell ${\bm r} = 0$, where the impurity is located, using the Dirac cone approximation.
If the full forms, Eqs. (\ref{imGbar}) and (\ref{reGbar}), of the Green's are used, 
an analytic close form is still possible, but is somewhat cumbersome.

\begin{figure}[ht]
\begin{center}
\includegraphics*[angle=0,width=8cm]{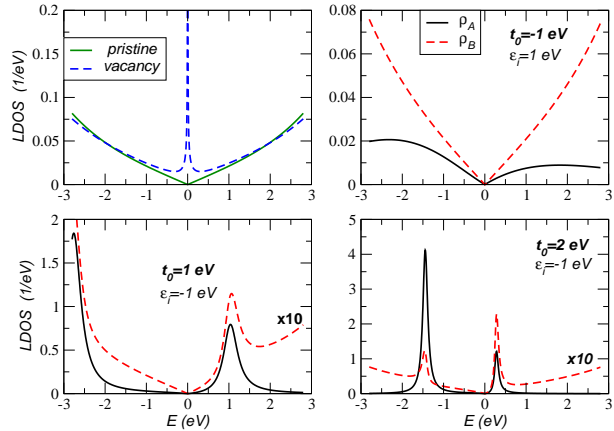}
\caption{Local density of states, $\rho_x(E)$, for $x=a,b$ at
the graphene's unit cell $\bm r=0$. We have used $t=3$ eV.
Upper left: density of states of pristine graphene and at the $B$ site
close to a vancancy. All other panels: local density of states at $A$ (impurity)
 and $B$ (next to the impurity)
sites for different values of the parameters $t_0$ and $\epsilon_i$.
The values of $t_0$ and $\epsilon_i$ corresponds to different types of impurities.
 \label{Fig:ldos_adatom}}
\end{center}
\end{figure}

For the calculation of the local density of states,
the case of a vacancy and the case where $\epsilon_i\ne0$ and $t_0\ne t$ have to be treated
separately. For the vacancy, the density of states at the neigbouring $B$ atom is
\begin{equation}  
\label{eq:LDOS1_Dirac}
  \rho_B (0,\omega) =
   \frac{2}{\sqrt 3\pi t } \abs{\frac{\hbar \omega}{t}}
   \left( 1 - \frac{1}{9} \abs{\frac{\hbar \omega}{t}}^2 + 
  \frac{1}{3} \abs{\frac{t}{\hbar \omega}}^2 L(\omega)
   \right) \,.
\end{equation}
with 
\begin{equation}
 L(\omega)=\left[ 1 + \frac{1}{\pi^2} \ln^2 \! \left( \frac{1}{\sqrt{3} \pi}
   \abs{\frac{\hbar \omega}{t}}^2 \right) \right]^{-1}\,.
\end{equation}
For the case of a general substituting atom, the local density of states 
at the impurity atom $\rho_a(0,\omega)$ is obtained from the
imaginary part of the Green's function which reads
\begin{eqnarray}  
 \Im G_{aa}(\omega) = \Im \bar{G}^0(\omega) \left\{ 1 + \frac{t_0 (2 t - t_0)(t-t_0)^2}{N_c |D(\omega)|^2} 
 \right.\nonumber\\
 - \frac{2 (t\!-\!t_0)^2 [\omega t_0(2 t\!-\!t_0)\!-\!\epsilon_i t^2]}{N_c |D(\omega)|^2} \Re \bar{G}^0(\omega)
\left. -\frac{[\omega t_0(2 t\!-\!t_0)\!-\!\epsilon_i t^2]}{N_c |D(\omega)|^2} [\Im \bar{G}^0(\omega)]^2
  \right\} .
\label{imGaa}
\end{eqnarray}
The imaginary part of the Green's function at the $B$ site, 
next nearest neighbor to the impurity, reads
\begin{eqnarray}
\Im G_{bb}(\omega) =  \Im \bar{G}^0(\omega) \left\{ 1 - \frac{t^2 [\omega t_0 (2t-t_0)-\epsilon_i t^2]}{9 N_c |D(\omega)|^2} \left( 1-\frac{\epsilon_i}{\omega}\right)\right\}\,.
\label{imGbb}
\end{eqnarray} 
If in Eqs. (\ref{imGaa}) and (\ref{imGbb}) one uses the full result for $\bar{G}^0(\omega)$,
given by Eqs. (\ref{imGbar}) and (\ref{reGbar}), the 
resulting expressions are valid for energies up to $2.5$ eV.
The local density of states $\rho_x(\bm r=0,\omega)$, $x=a,b$, are depicted in
Fig. \ref{Fig:ldos_adatom}, for different choices of $t_0$ and $\epsilon_i$.

It is well-known that the density of states due to a vacancy has a strong departure
from the pristine value close to the Dirac point. This is due to the
logarithmic singularity seen in Eq. (\ref{eq:LDOS1_Dirac}). In the
case where there is an enhancement of the hopping amplitude between the impurity and the
neighboring atoms (negative $t_0$) the local density of states retains
its linear behavior close to the Dirac point, but its value at the
$A$ and $B$ sublattices are different, as expected (Fig. \ref{Fig:ldos_adatom}, upper right panel). This case would mimic
a boron impurity atom.
Boron has a larger atomic radius ($R \simeq 0.85$ \AA) than carbon ($R \simeq 0.7$ \AA), and there should be an increase in the absolute value of the hopping amplitude when it substitutes a carbon in graphene. 
In the case of a decreasing of the electron hopping between the impurity and the
carbon atoms (positive $t_0$) the behavior of the density of states
is more interesting since resonances
start to develop around the Dirac point (Fig. \ref{Fig:ldos_adatom}, two lower panels). Note that the density of states
still goes to zero
at the Dirac point. This behaviour is reminiscent of the vacancy, since we
can picture the two resonances developed in both sides of the Dirac point as a
splitting of the divergent peak for the vacancy due to the departure of $t_0$
from its vacancy value $t_0=t$. This case would mimic a nitrogen atom, which has a smaller atomic radius ($R \simeq 0.65$ \AA) than carbon.

The calculation of the local density of states for finite $\bm r$ requires the
calculation of the Fourier transform entering the definition in Eq. 
(\ref{realspaceG}). For $G_{bb}(\bm r,\bm r,\omega)$ we use an approximation
for $\phi(\bm k)$, which reads $\phi(\bm k)\simeq 3a_0(k_y-ik_x)/2$.
Carrying out the Fourier transform we obtain
\begin{eqnarray}
G_{aa}(\bm r,\bm r,\omega)=\bar G^0(\omega) + T(\omega)[F_0(\omega,r)]^2\,,\\
G_{bb}(\bm r,\bm r,\omega)=\bar G^0(\omega) + \frac {t^2}{\omega^2}T(\omega)[F_1(\omega,r)]^2\,,
\end{eqnarray}
where $F_n(\omega,r)$ ($n=0,1$) is defined as
\begin{equation}
F_n(\omega,r)=\frac{(2n+1)A_ca_0^n}{2^{n+1}\pi v_F^2} 
\left[
\frac {2\omega}{r^n}\dashint_0^{k_cr}dx\frac{x^{n+1}J_n(x)}{\alpha^2-x^2}
-i\frac{\pi}{v_F^n}J_n(\alpha)\abs{\omega}^{n+1}
\right]\,,
\label{Fn}
\end{equation}
with $J_n(x)$ the Bessel function of integer order $n$; and
 $k_c=2\sqrt\pi/(\sqrt{3\sqrt{3}}a_0)$, $\alpha = \abs{\omega} r /v_F$,
$A_c=3\sqrt 3 a_0^2/2$, and $v_F=3ta_0/2$.  The Cauchy principal value
of the integral in Eq. (\ref{Fn}) is computed
using numerical methods.

\begin{figure}[ht]
\begin{center}
\includegraphics*[angle=0,width=8cm]{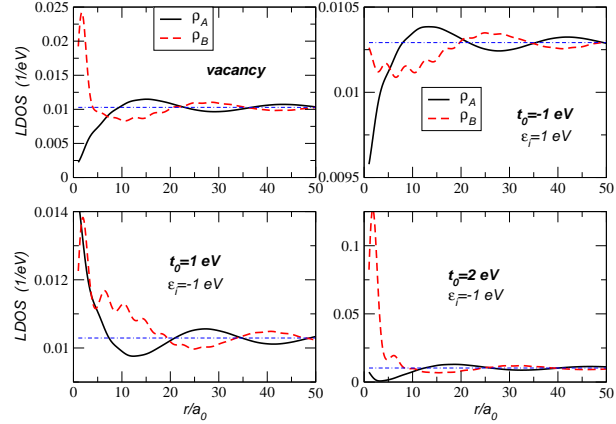}
\caption{Local density of states in sublattice $A$ and $B$
as function of $r>0$. The dashed-dotted line is the density of states
of pristine graphene at the energy $\hbar\omega=0.5$ eV. The
point $r=0$ is excluded. The small wavelength oscillations are due to the
cut-off momentum $k_c$ and the large ones are the $2q_F$ Friedel oscillations.
 \label{Fig:ldos_R}}
\end{center}
\end{figure}
In Fig. {\ref{Fig:ldos_R}}, we plot the local density of states at
both sub-lattices $A$ and $B$. The typical oscillations due to the
presence of the impurity are present. Note that close to the impurity
the $B$ sublattice density of states presents higher harmonics
as function of $r$; these are due to the cut-off momentum $k_c$. 
On the other hand, the large wavelength oscillations are the $2q_F$
Friedel oscillations: from Fig. \ref{Fig:ldos_R} the wavelength is about
$\lambda\simeq 28 a_0$; on the other hand, for the energy $\hbar\omega=0.5$ eV the Fermi
momentum is $q_F=1/(9a_0)$, implying $\lambda\simeq \pi/q_F\simeq 28a_0$.
At large values of $r$ the two density of states
are out of phase by a factor of $\pi$. Therefore, when we average over
the unit cell the result is essentially the pristine density of states.
This result has strong consequences for STM experiments. If the STM 
experiment lacks atomic resolution at the $A$ and $B$ sublattices level,
the experimental data will show a very faint trace of the $2q_F$ Friedel oscillations.
This seems to be the case in the experiments reported in Ref. \cite{Bose}.
Closer to the impurity there is a strong departure from the asymptotic
behavior \cite{bena}
\begin{equation}
 \rho(\bm r)\propto \frac{1}{r^2}\sin(2r\omega/v_F)\,,
\end{equation}
and the $A$ and $B$ LDOS behave quite differently.

STM measurements of a material surface
are ideal for studying real-space local features with atomic resolution. 
In particular, real space modulations of the STM intensity can be observed when impurities are
present at the surface of a given material. 
In general, the impurities lead to elastic scattering between the momentum $\bm q_F$
and $-\bm q_F$, which is the most efficient process due to phase space restrictions, leading to $2q_F$ Friedel oscillations.
In the case of graphene, the chiral nature of its electronic spectrum changes this
general behavior. The Fermi surface has disconnected pieces at different points
of the Brillouin zone -- the $\bm K$ and $\bm K'$ points. The Fermi surface consists of  circumferences
of radius $\bm q_F$ around each $\bm K$ and $\bm K'$ points. The scattering process
is then characterized by two channels: an intra-cone scattering 
(within the same $\bm K$ or $\bm K'$ points) of momentum change
$2\bm q_F$, and an inter-cone scattering (between the $\bm K$ and $\bm K'$) points.

A Fourier transform of the real-space STM-intensity currents, proportional
to the LDOS, will produce bright spots at the momentum values
 seen in the real space modulations of the LDOS. 
When impurities are present, the momentum values
characterizing the real space modulation are related to the momentum change
associated with a given scattering process.
In Ref. \cite{Bose} it was found that the intra-cone scattering, which would give rise
to a bright spot of radius $2q_F$, was absent in the momentum map of the density of states obtained by a Fourier transform of their STM data.

\begin{figure}
\begin{center}
\subfigure{\includegraphics[width=0.3\textwidth,angle=270]{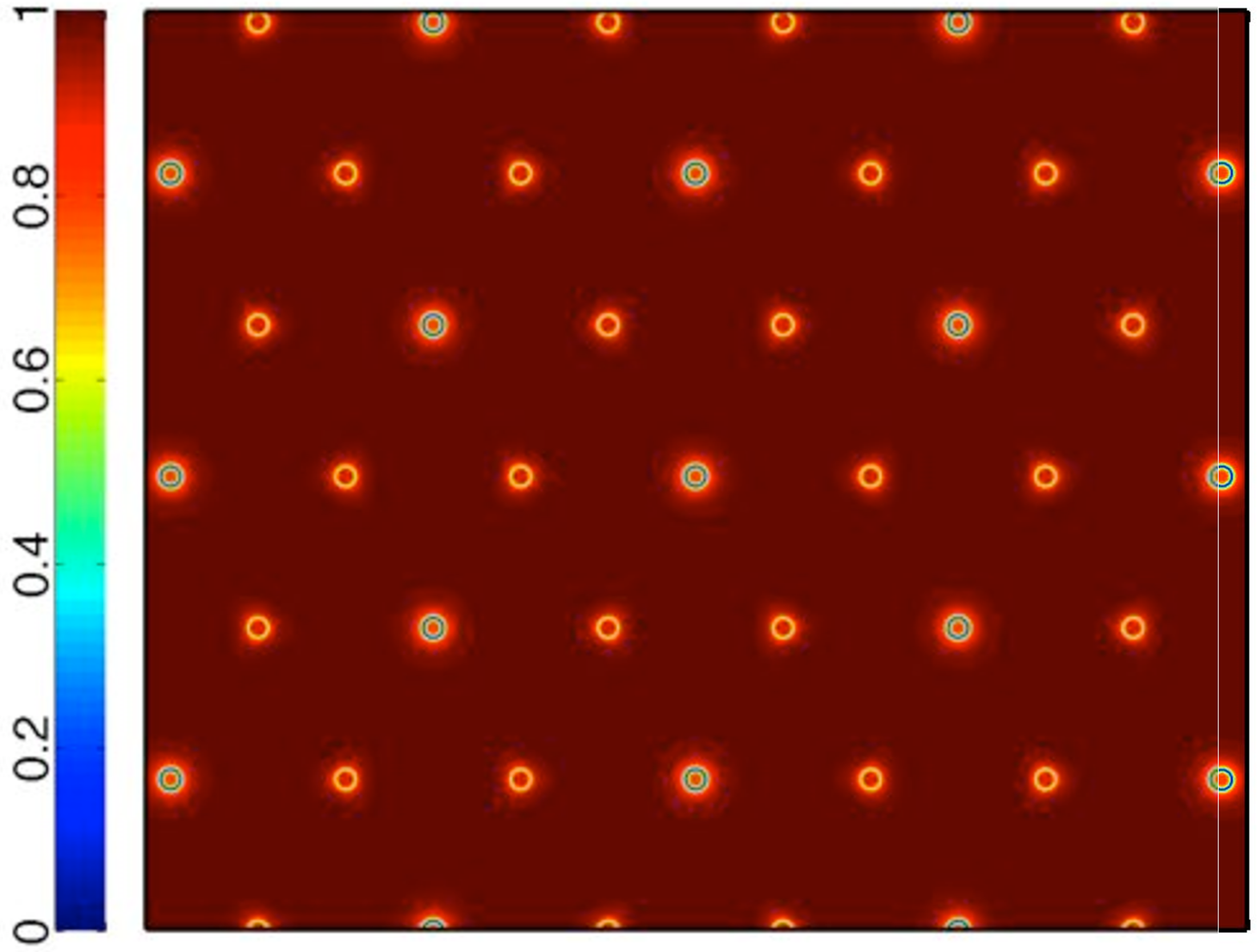}}
\hfil
\subfigure{\includegraphics[width=0.3\textwidth,angle=270]{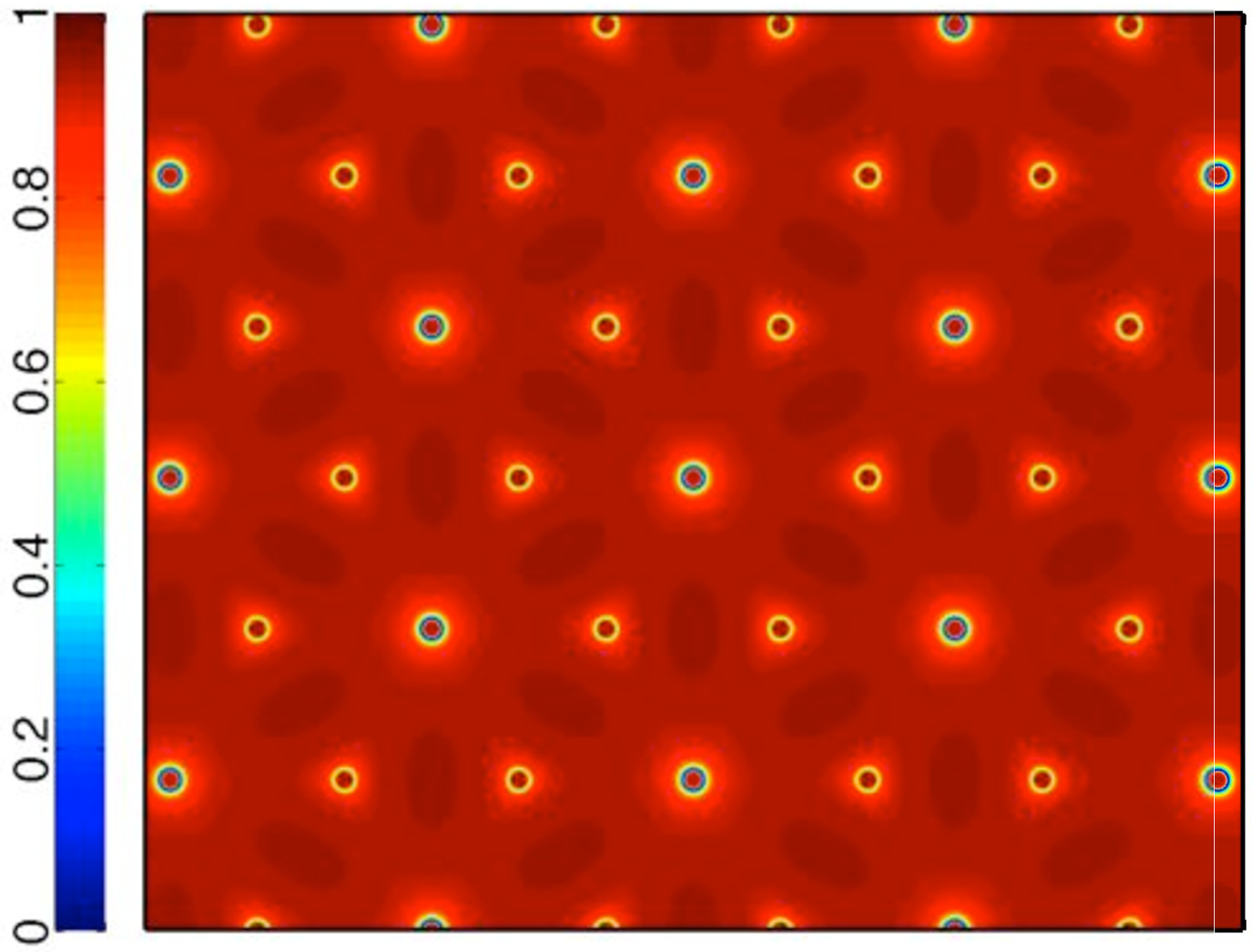}}
\hfil
\subfigure{\includegraphics[width=0.3\textwidth,angle=270]{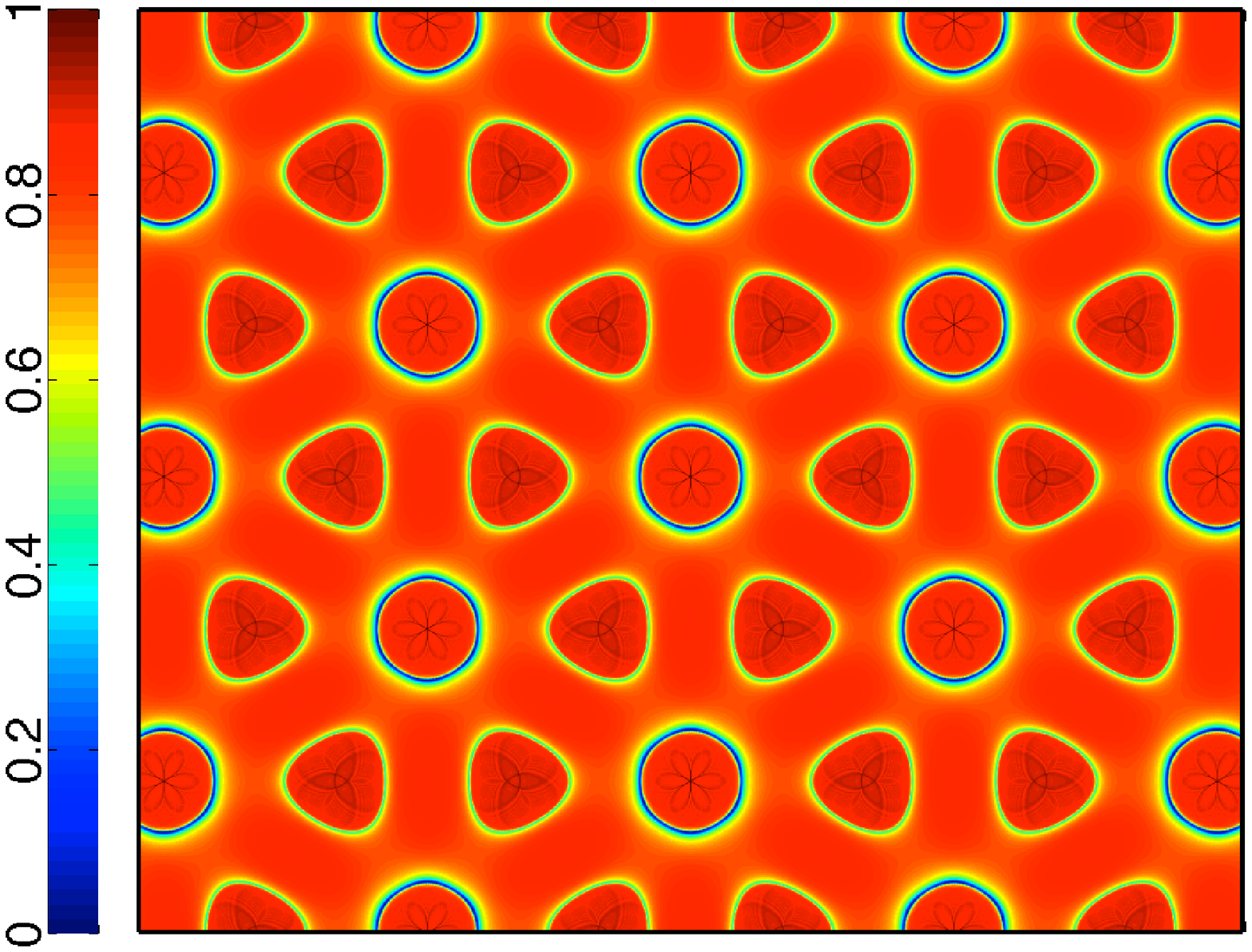}}
\hfil
\subfigure{\includegraphics[width=0.3\textwidth,angle=270]
{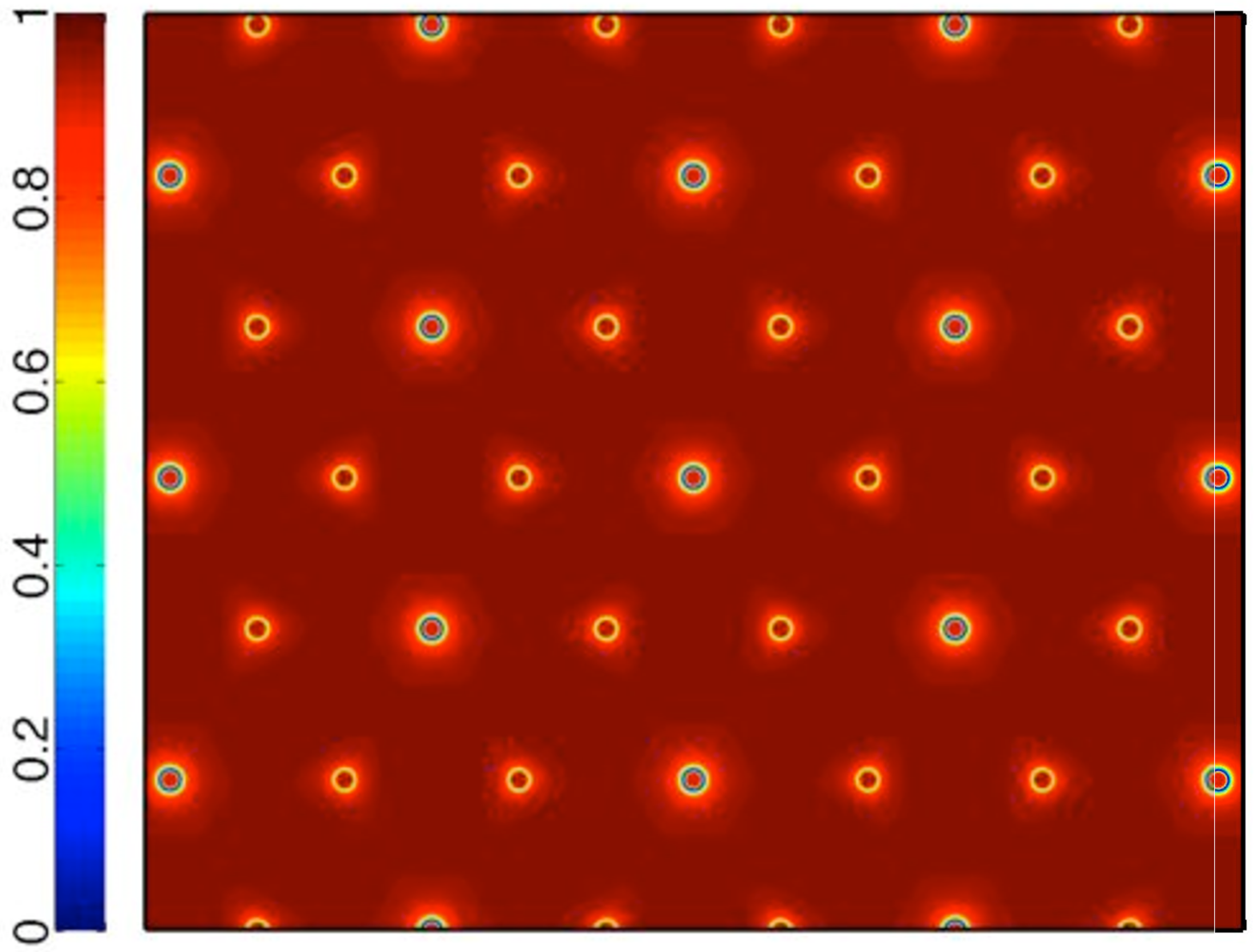}}
\subfigure{\includegraphics[width=0.28\textwidth,angle=270]{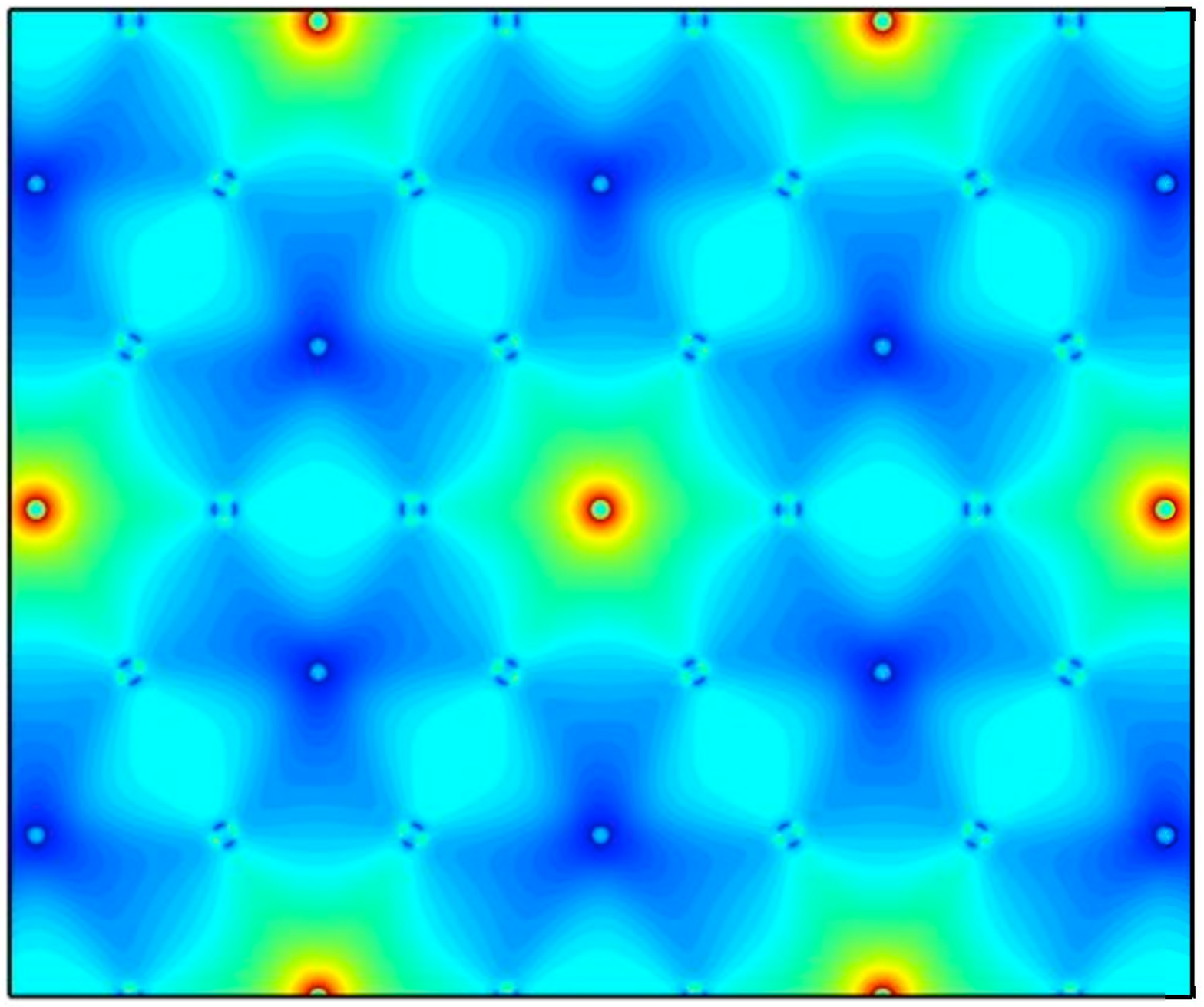}}
\hfil
\subfigure{\includegraphics[width=0.28\textwidth,angle=270]{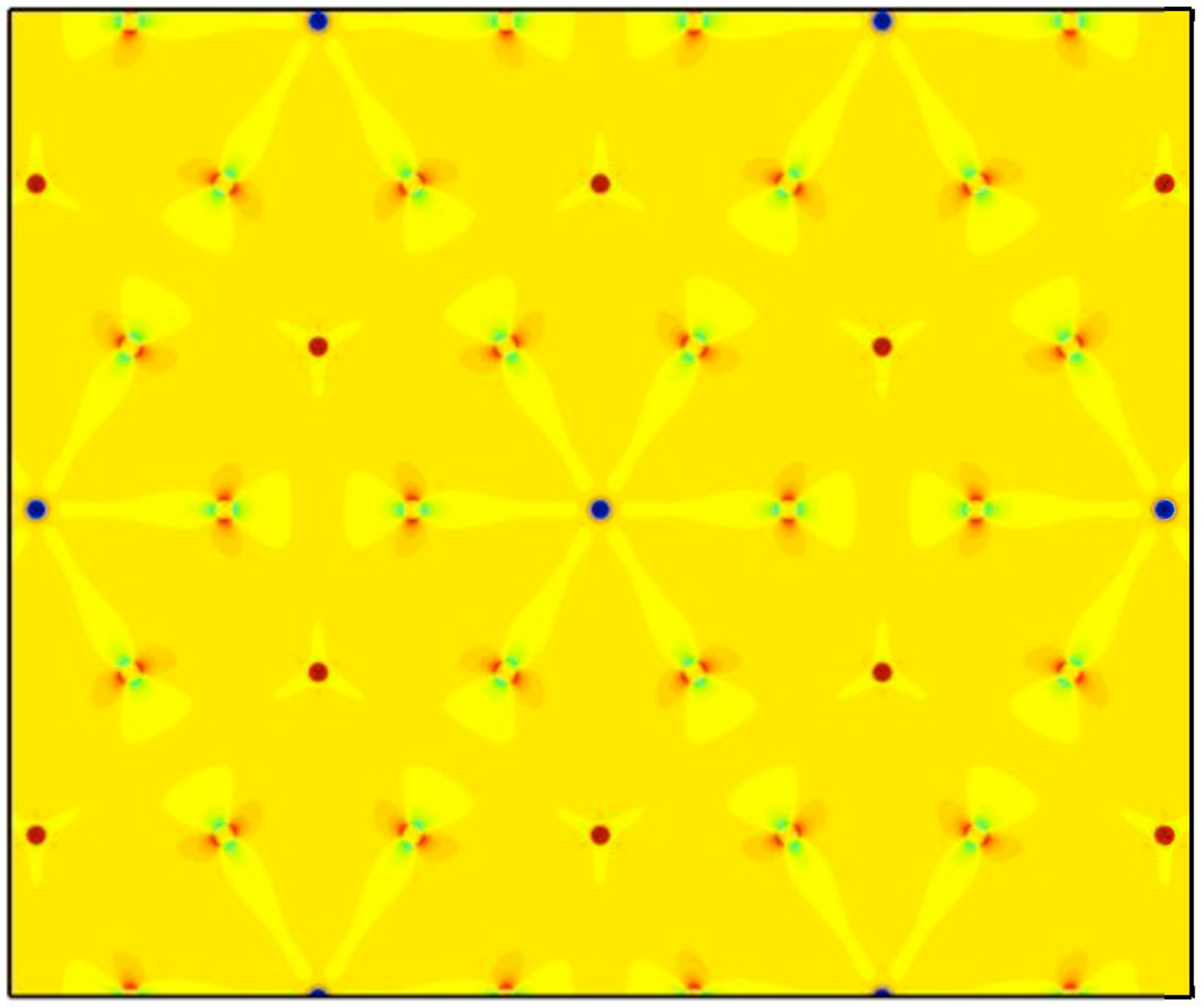}}
\hfil
\subfigure{\includegraphics[width=0.28\textwidth,angle=270]{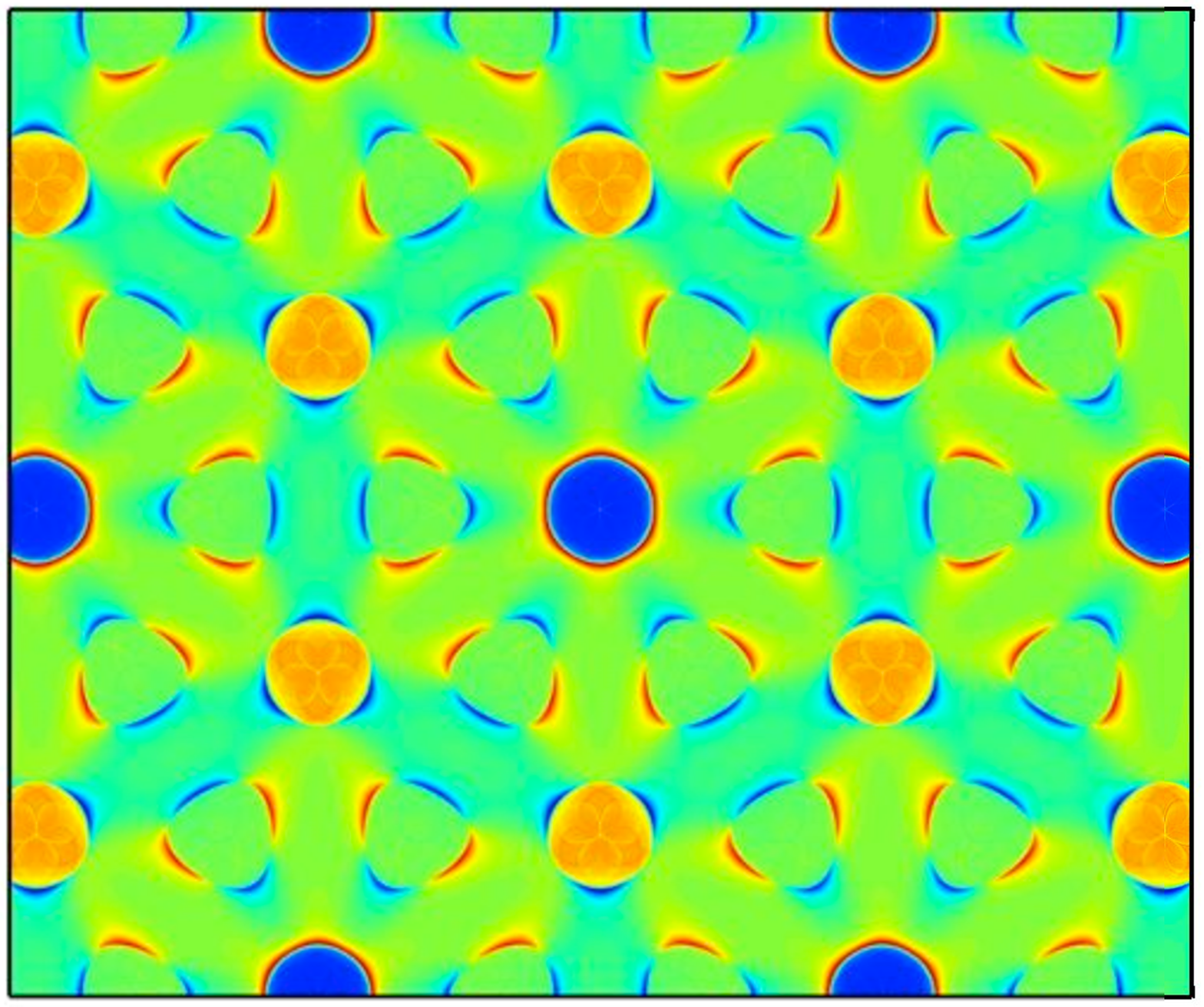}}
\hfil
\subfigure{\includegraphics[width=0.28\textwidth,angle=270]{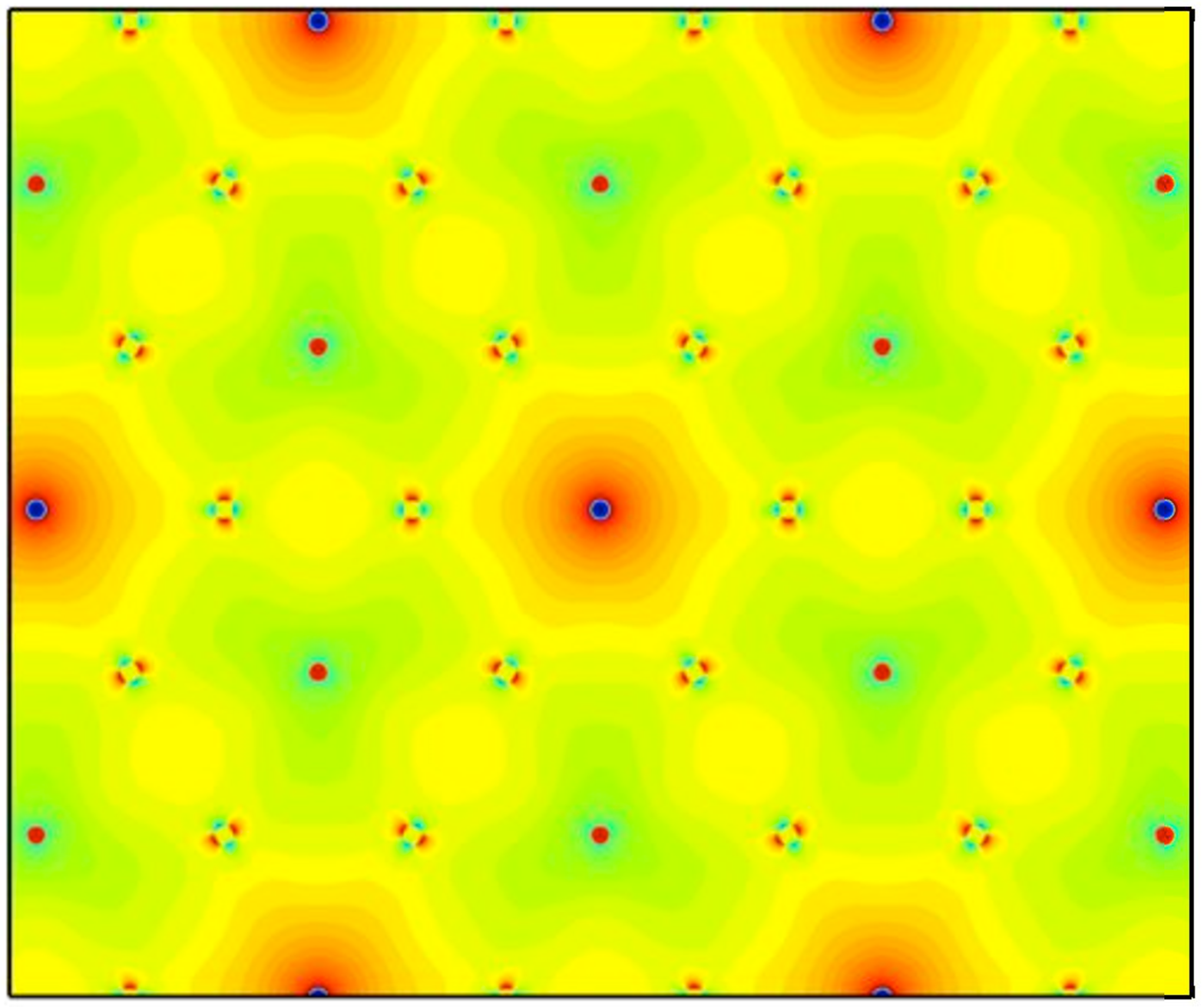}}
\subfigure[a)]{\includegraphics[width=0.28\textwidth,angle=270]{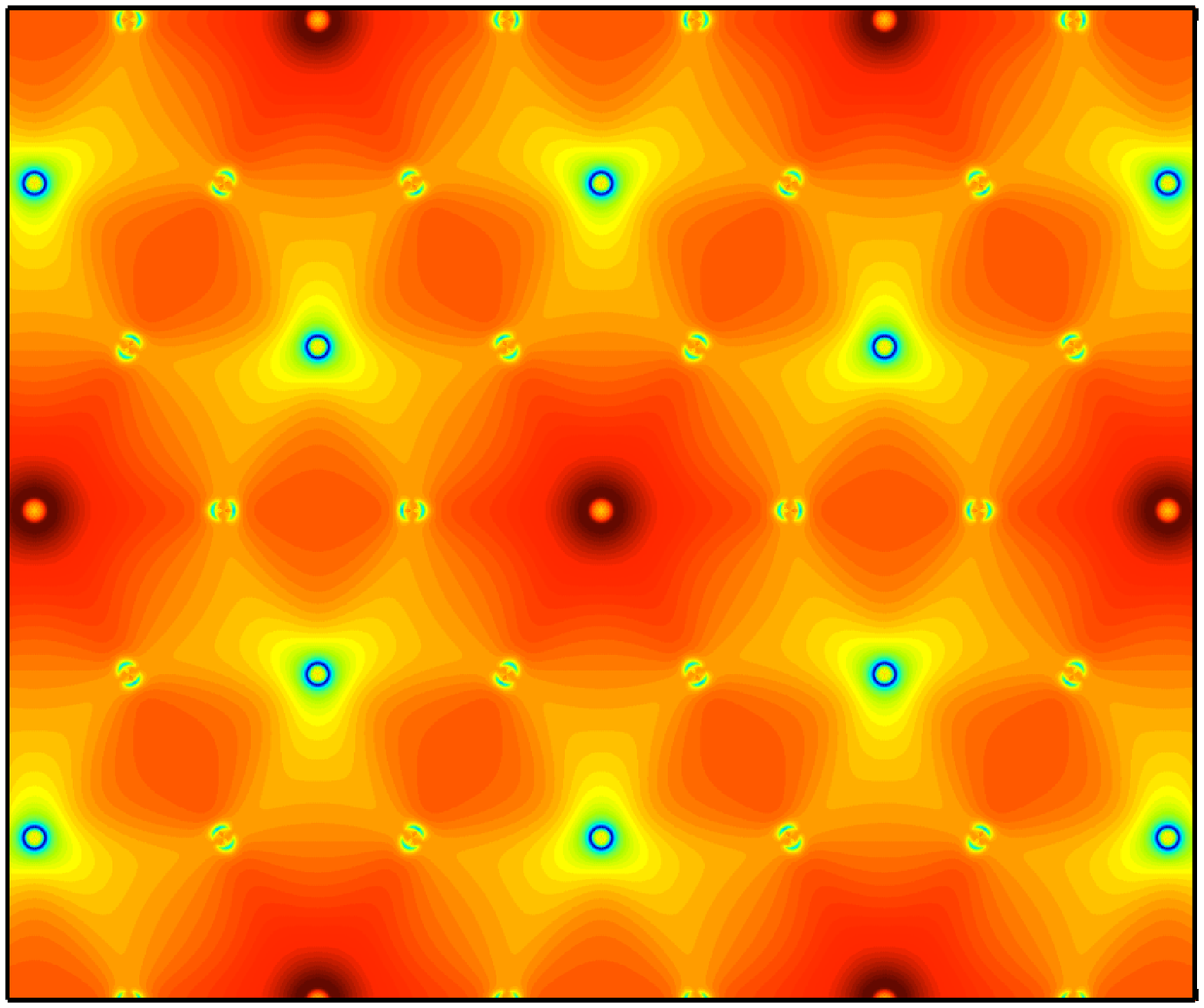}}
\hfil 
\subfigure[b)]{\includegraphics[width=0.28\textwidth,angle=270]{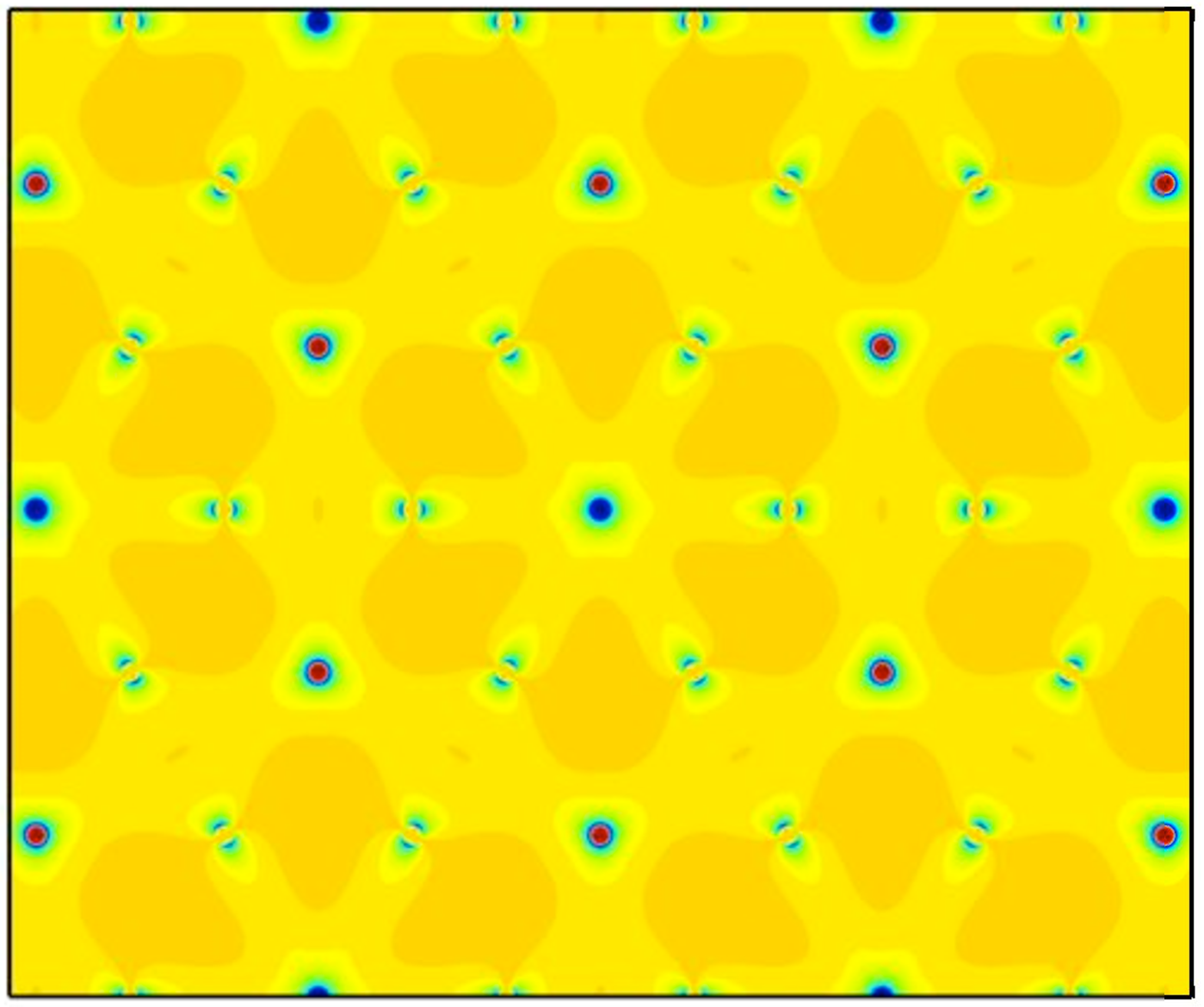}}
\hfil 
\subfigure[c)]{\includegraphics[width=0.28\textwidth,angle=270]{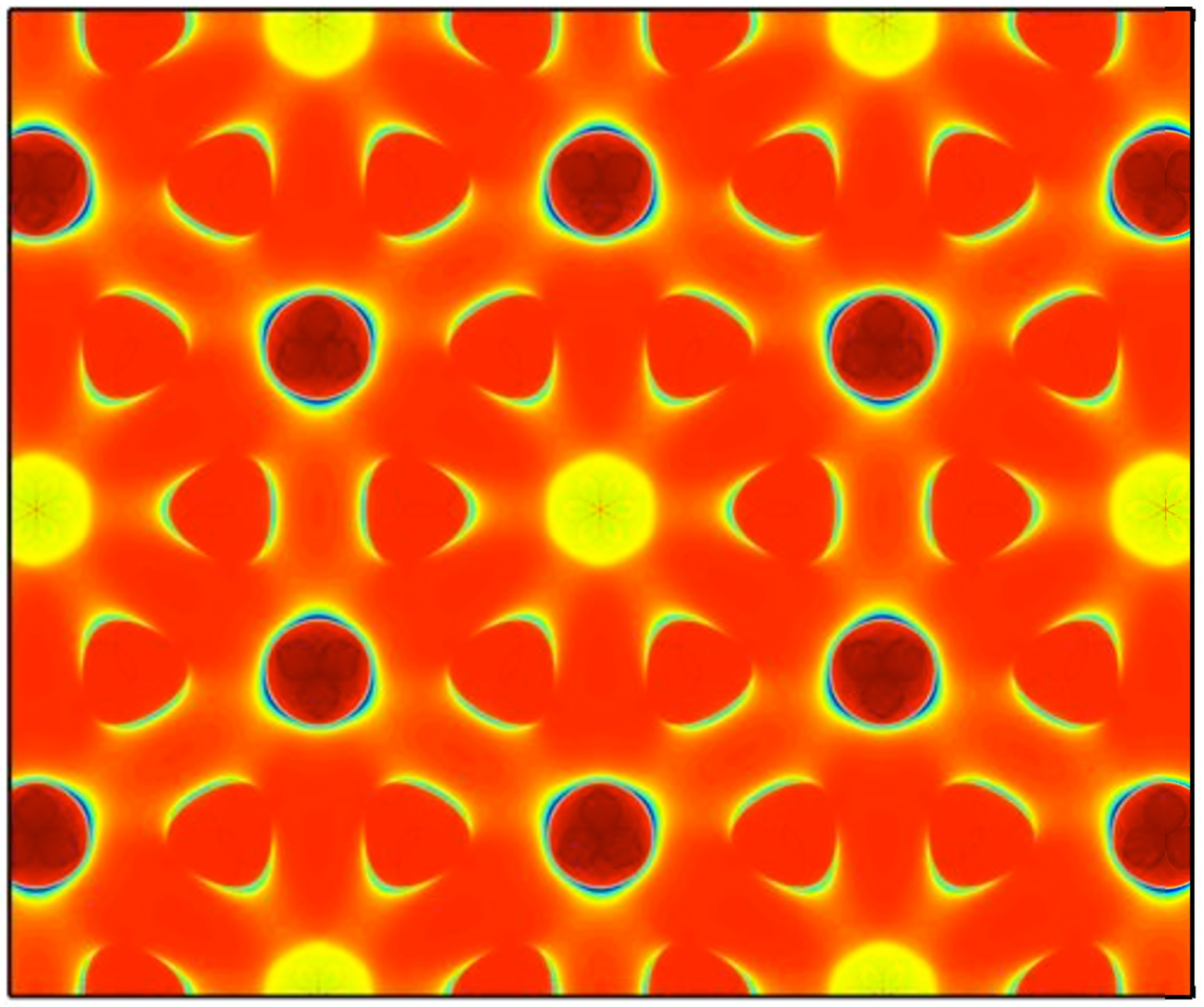}}
\hfil 
\subfigure[d)]{\includegraphics[width=0.28\textwidth,angle=270]{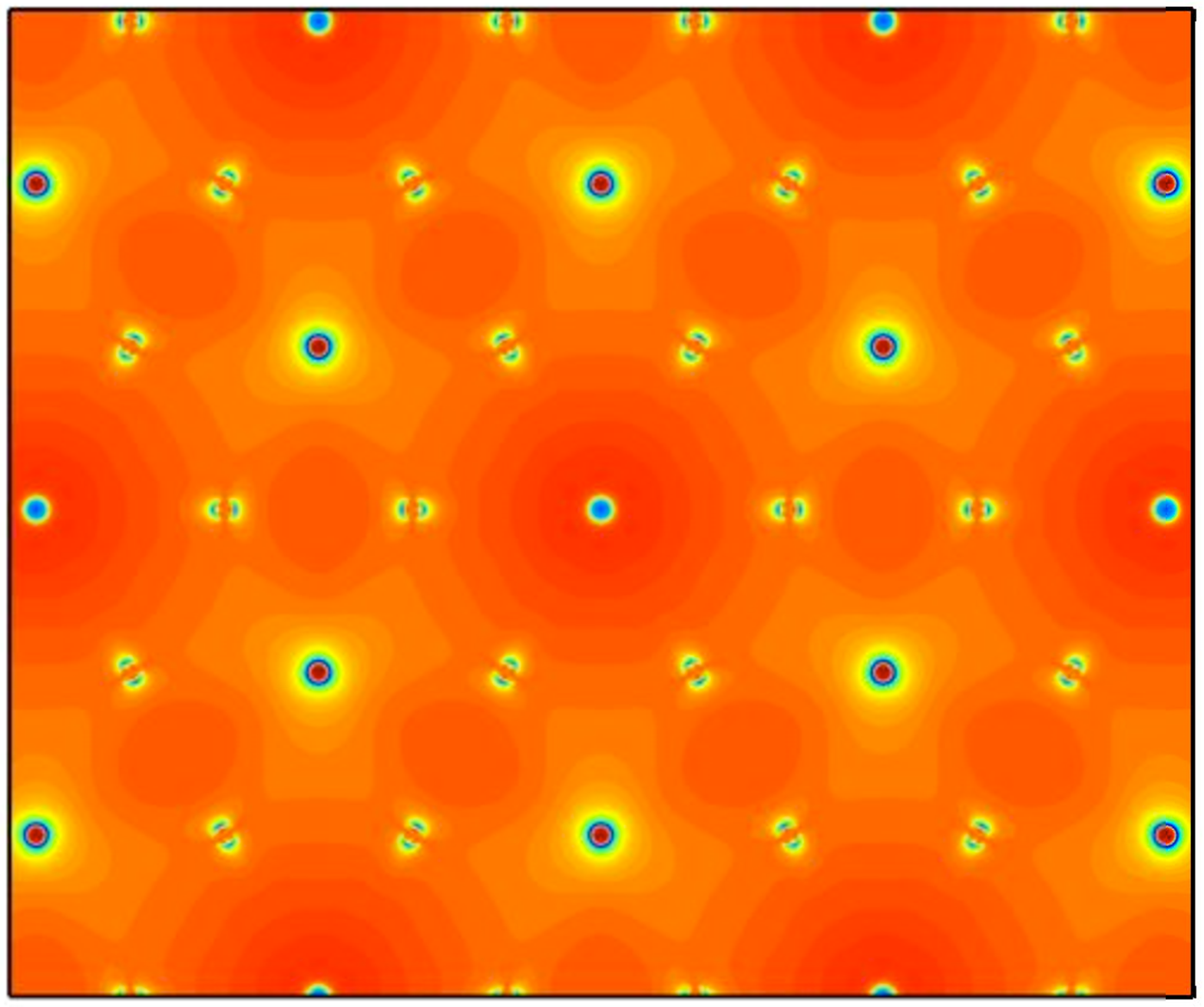}} 
\end{center}
\caption{Fourier transform of the local density of states. The first row is the density of states in sublattice A ($\rho_a$), the second row is the density of states in sublattice B ($\rho_b$), and the third row is the sum of the two, $\rho_a+\rho_b$. Column (a) is for the case of a vacancy, and $\omega=0.3$ eV. Column (b): $t_0=-1$ eV, $\epsilon_i=1$ eV, $\omega=0.3$ eV. Column (c): $t_0=1$ eV, $\epsilon_i=-1$ eV, $\omega=-1.5$ eV. Column (d): $t_0=2$ eV, $\epsilon_i=-1$ eV, $\omega=-0.3$ eV.}
\label{Fig:rho_k}
\end{figure}

In what follows, we give Fourier transforms of the local density of states of graphene for 
the different types of impurities discussed previously in the text. We note that our derivation
of the Fourier transform of LDOS uses the full Green's functions for the calculation of the LDOS,
and therefore no approximation has been made in the calculation.
The full real-space map of density of states can be obtained numerically from Eqs. (\ref{rho_x}) and (\ref{realspaceG}), using the exact expressions Eqs. (\ref{gaa})-(\ref{green}), as was done in Ref. \cite{Peres2007}. We now perform a Fourier transform of the density of states,
\begin{eqnarray}
\rho_{x}(\bm k,\omega) &=& 
\sum_{\bm r}e^{-i\bm k\cdot \bm r}
 \rho_{x}(\bm r, \omega)\ ,
\label{kspacerho}
\end{eqnarray}
for each sublattice $x=a,b$. The results are shown in Fig. \ref{Fig:rho_k}, where the three rows correspond to $\rho_a(\bm k,\omega)$, $\rho_b(\bm k,\omega)$, and the sum $\rho_a(\bm k,\omega)+\rho_b(\bm k,\omega)$, from top to bottom; and the four columns correspond to four types of impurities discussed in Fig. \ref{Fig:ldos_adatom}.
We recall that positive $t_0$ reduces the hopping from the impurity site to its nearest neighbor carbon atoms -- the particular case of $t_0=t$ represents a vacancy -- and negative
$t_0$ increases the hopping of the electrons from the carbon atoms to the impurity
site. This latter case would correspond to an atom with a radius larger than carbon,
such as boron, leading to an increase of the hopping relatively to the hopping $t$ between nearest neighbor carbons.
\begin{figure}
\begin{center}
\subfigure{\includegraphics[width=0.3\textwidth,angle=270]{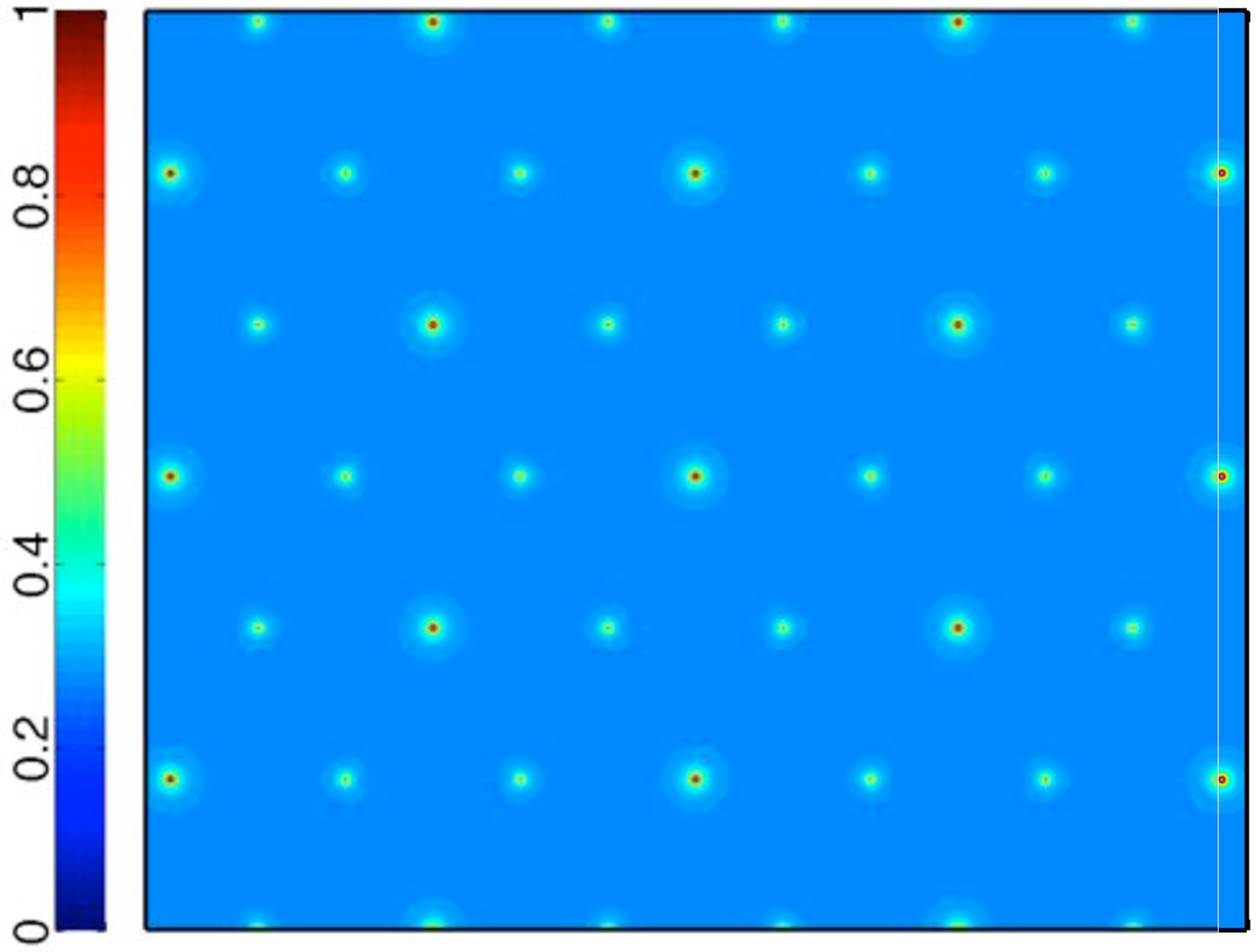}}
\hfil
\subfigure{\includegraphics[width=0.3\textwidth,angle=270]{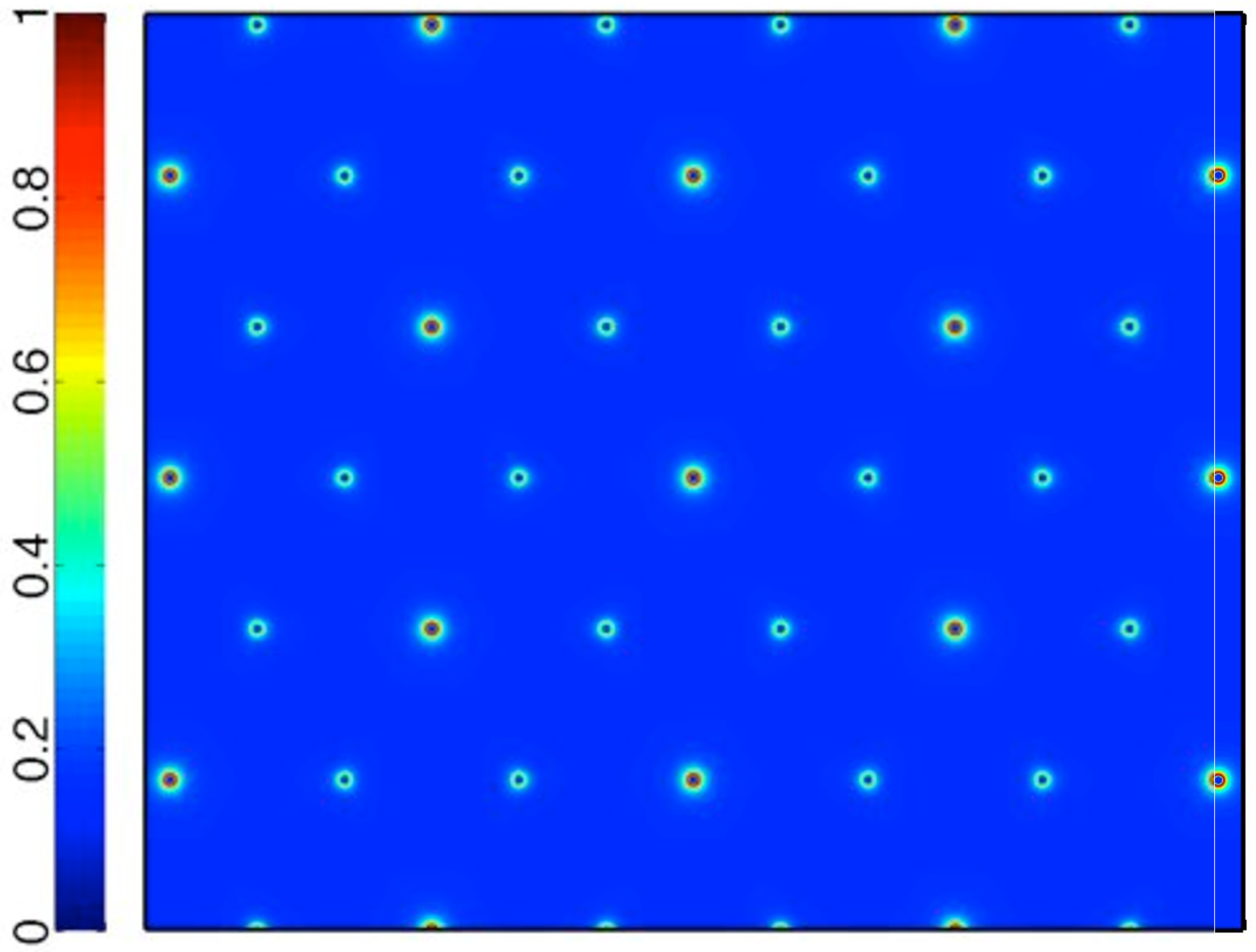}}
\hfil
\subfigure{\includegraphics[width=0.3\textwidth,angle=270]{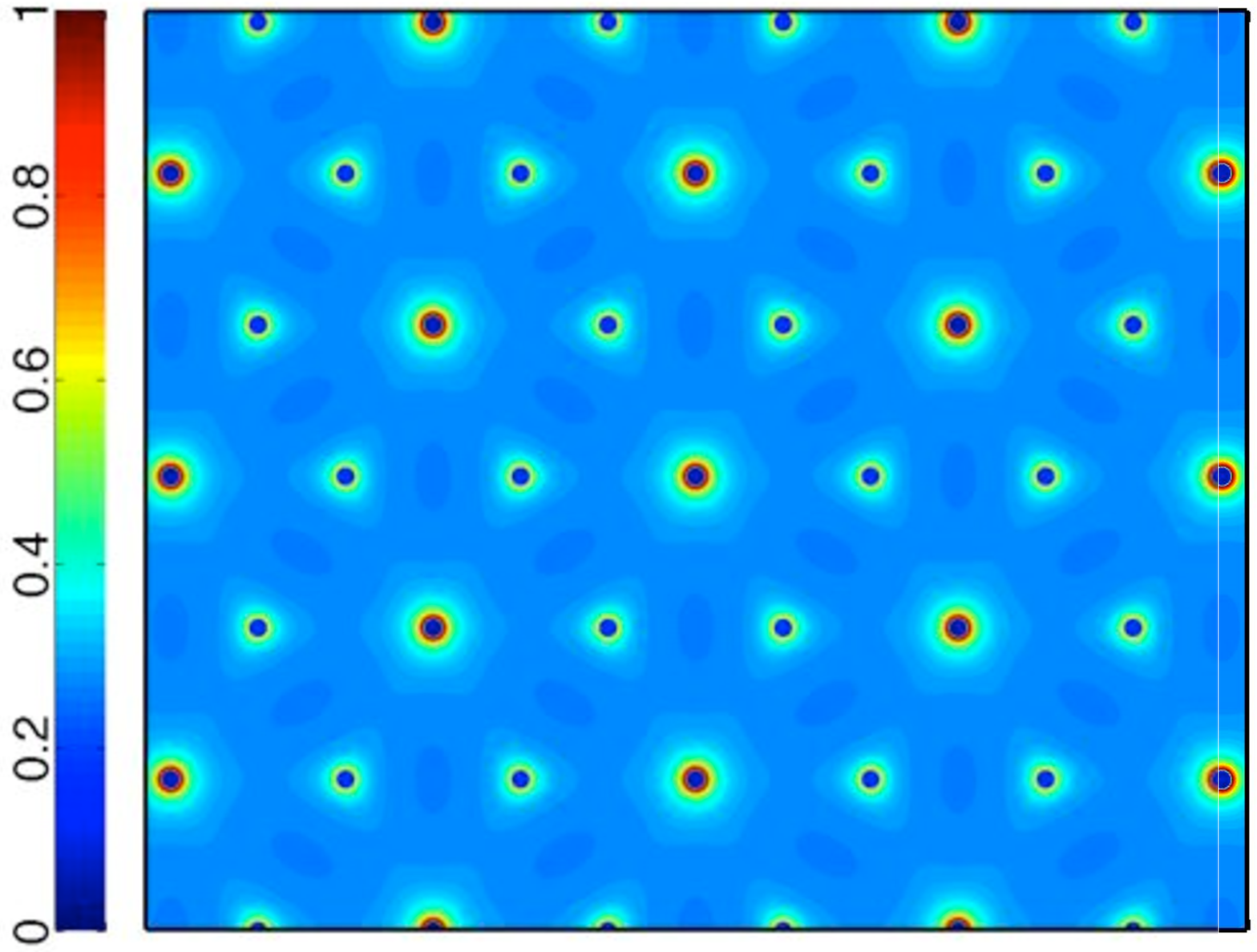}}
\hfil
\subfigure{\includegraphics[width=0.3\textwidth,angle=270]
{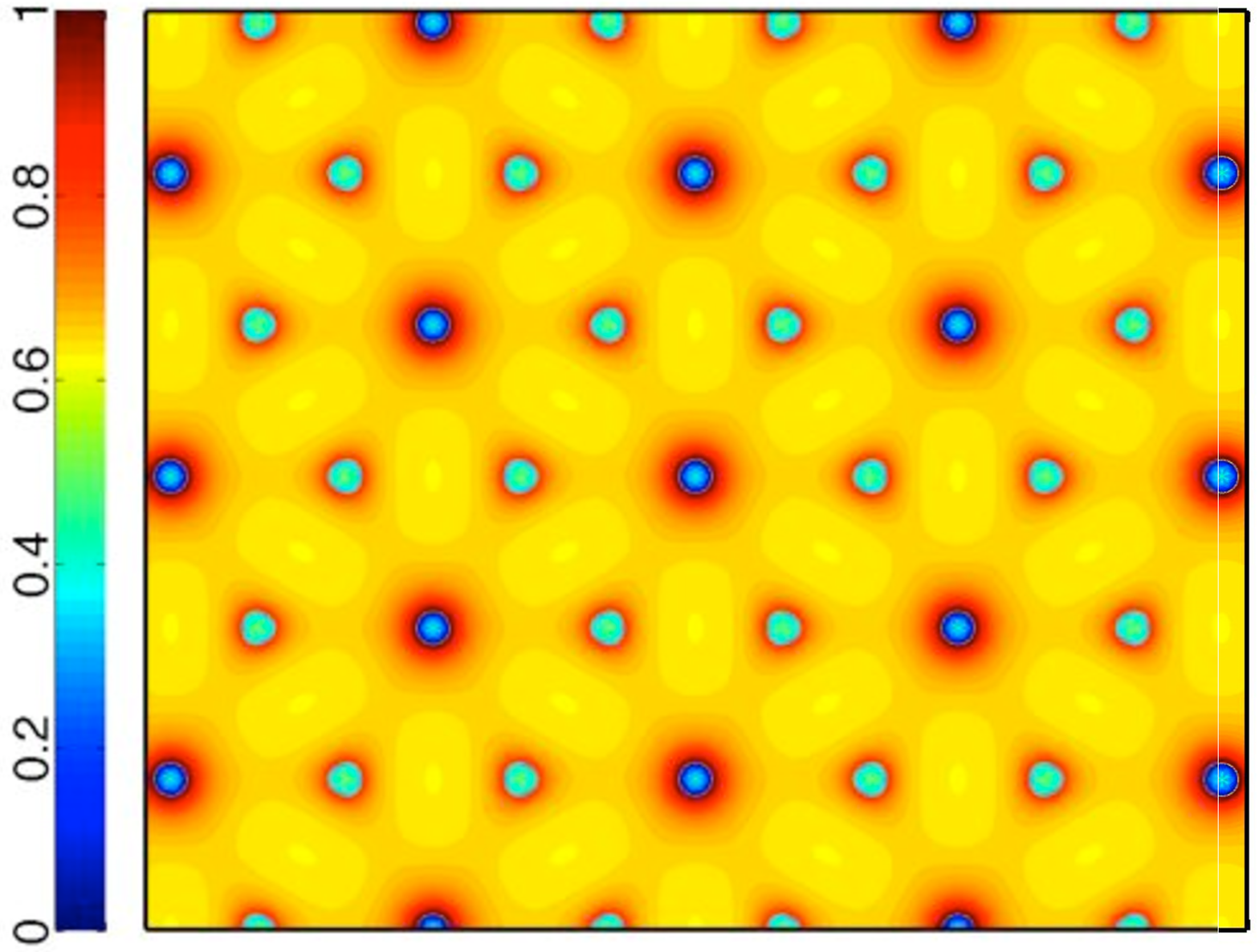}}
\subfigure{\includegraphics[width=0.28\textwidth,angle=270]{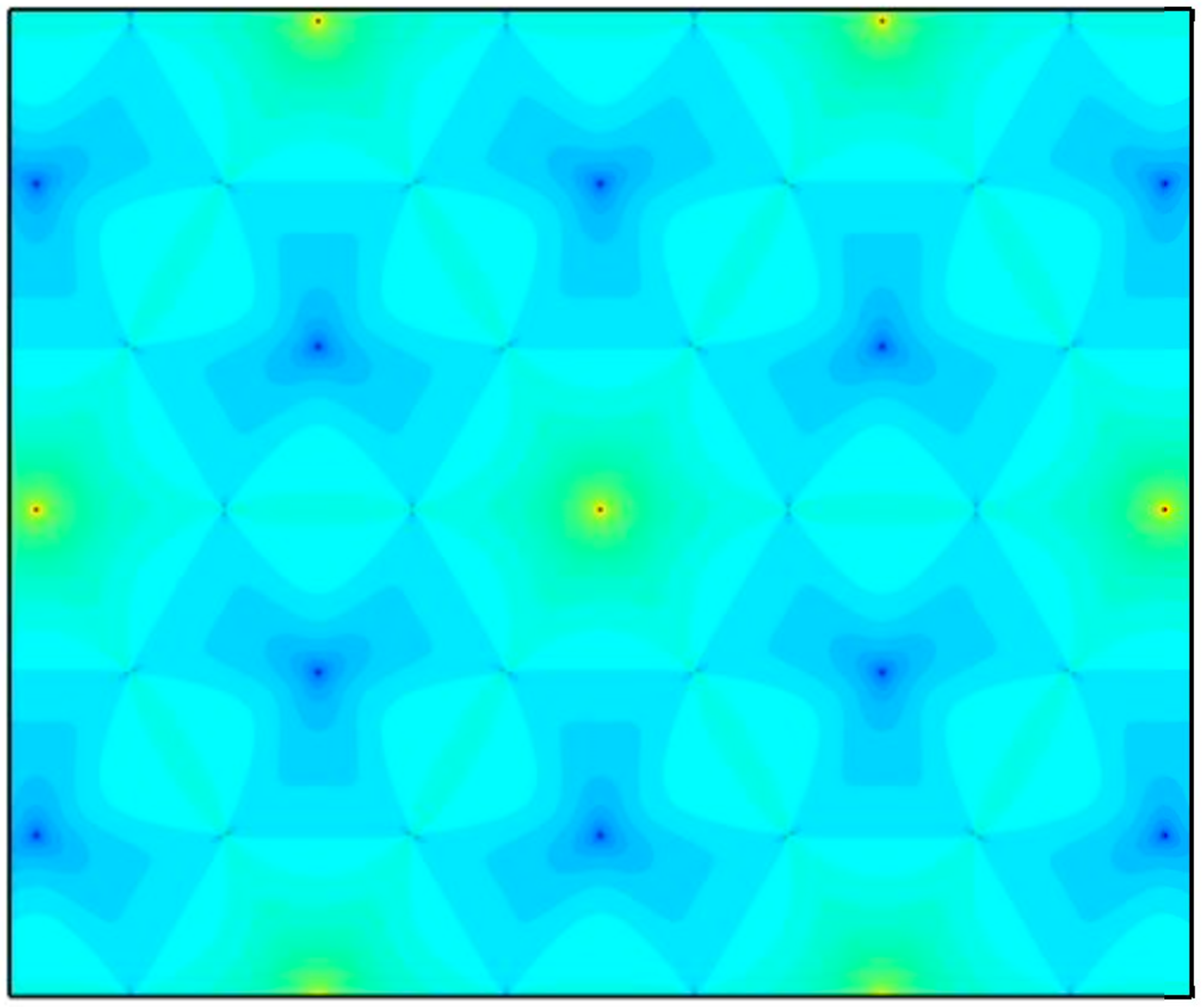}}
\hfil
\subfigure{\includegraphics[width=0.28\textwidth,angle=270]{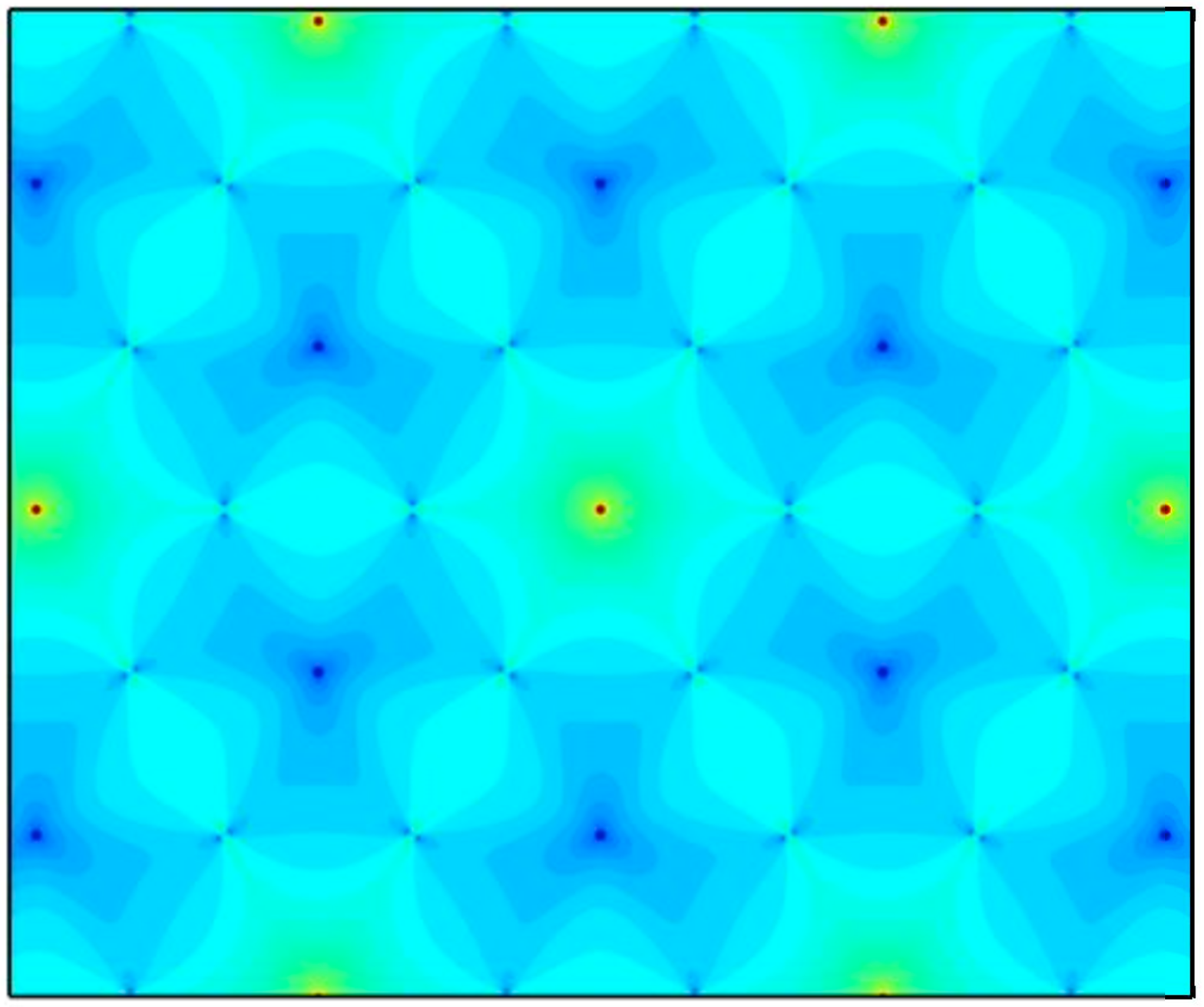}}
\hfil
\subfigure{\includegraphics[width=0.28\textwidth,angle=270]{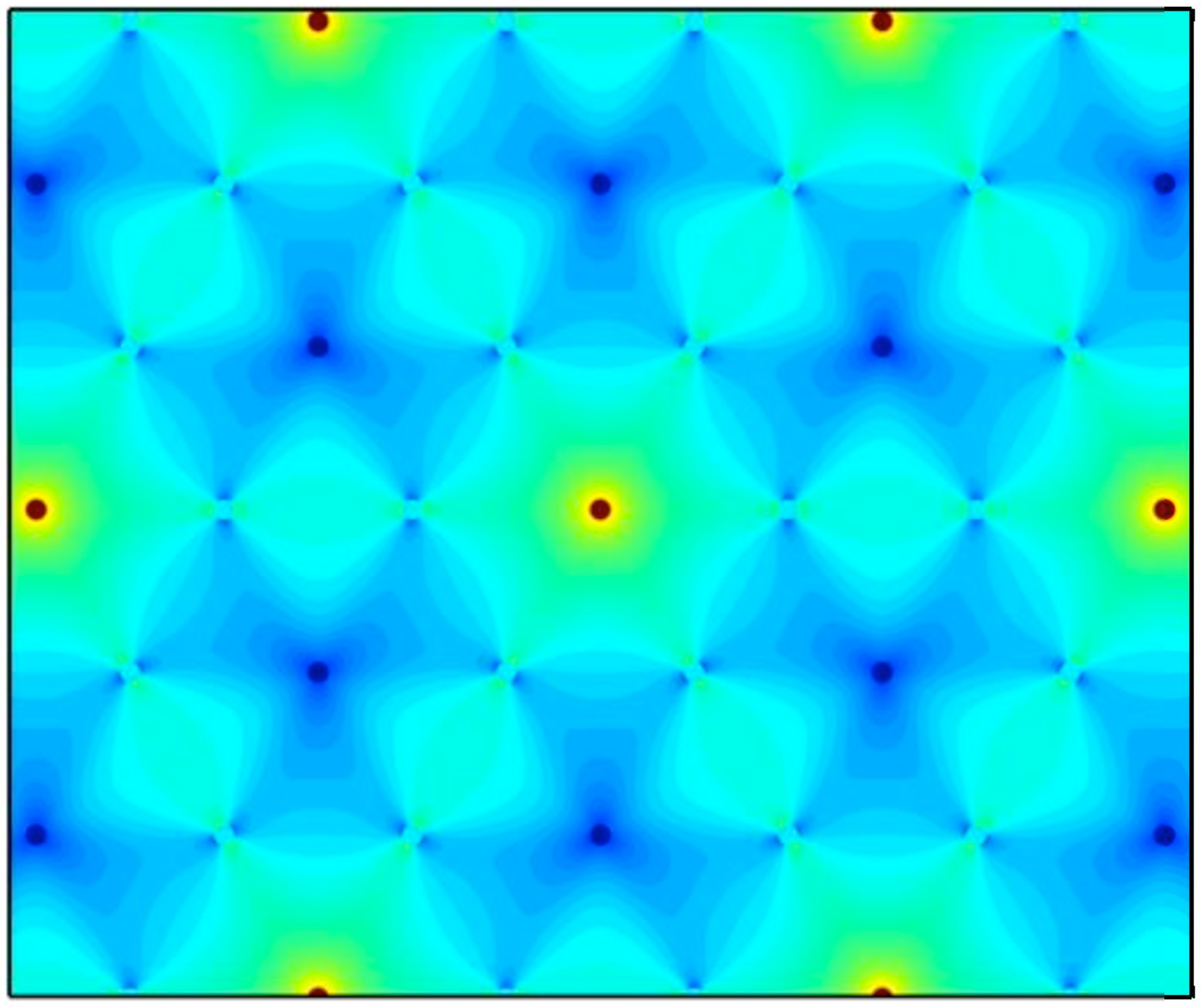}}
\hfil
\subfigure{\includegraphics[width=0.28\textwidth,angle=270]{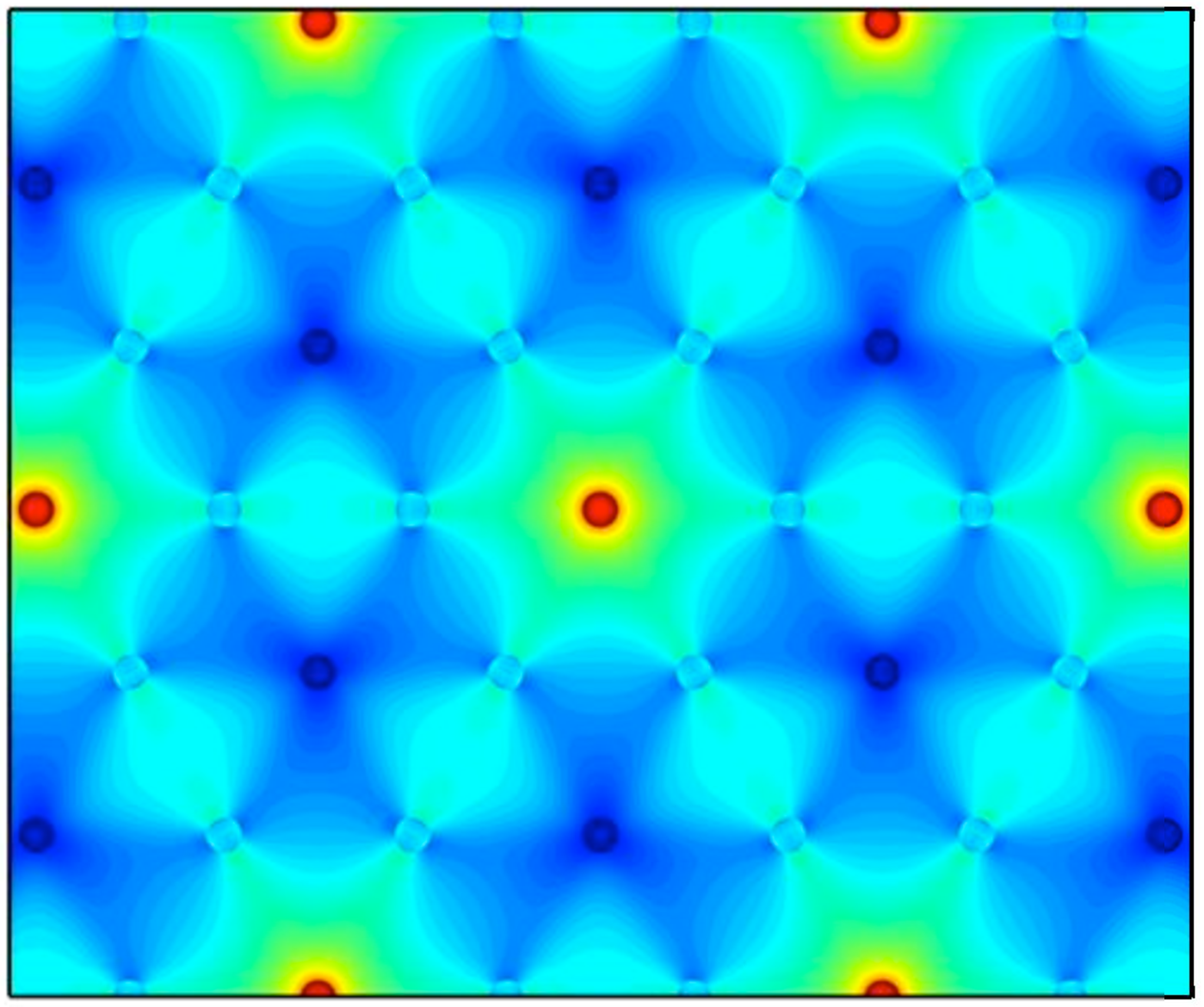}}
\subfigure[a)]{\includegraphics[width=0.28\textwidth,angle=270]{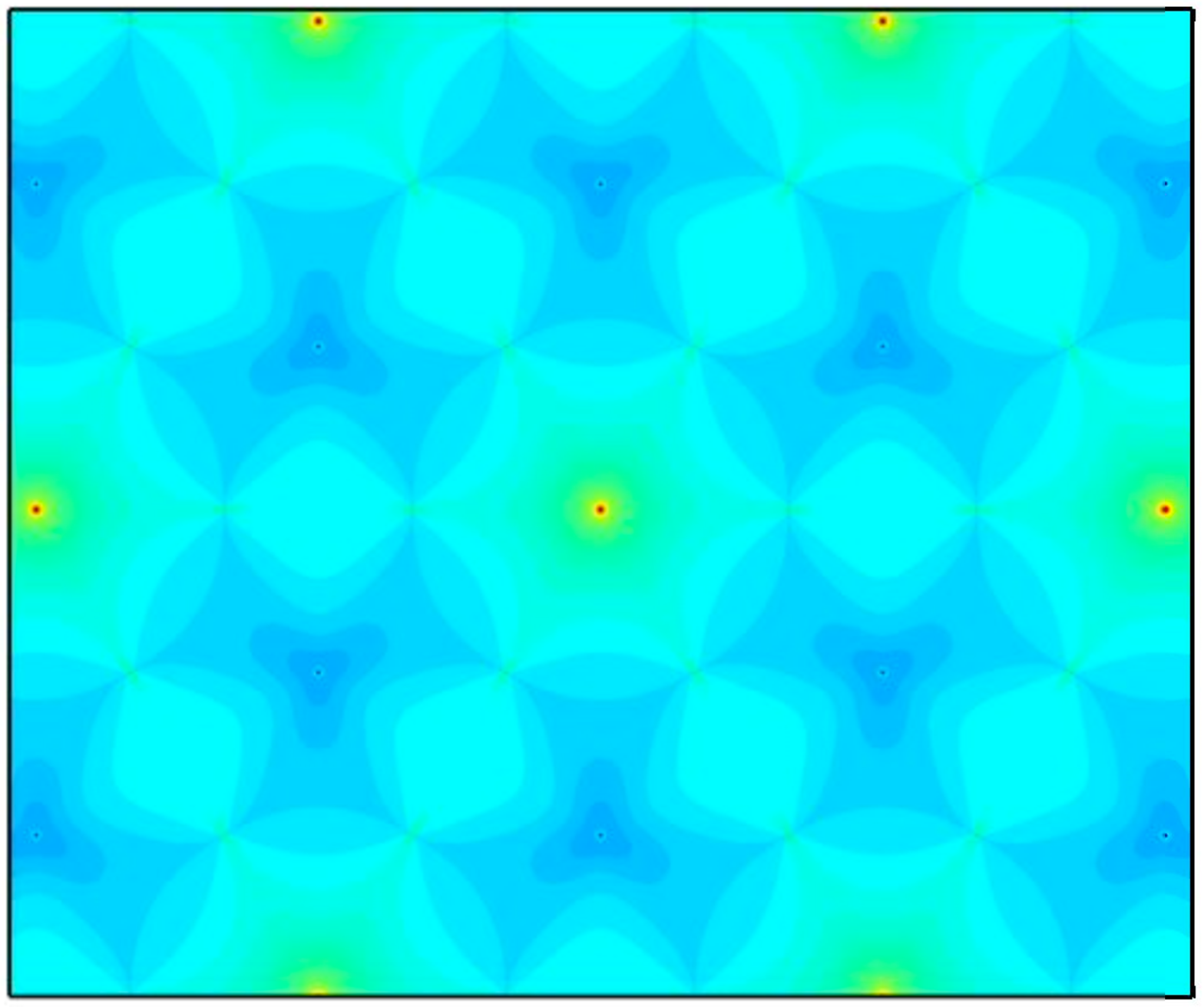}}
\hfil 
\subfigure[b)]{\includegraphics[width=0.28\textwidth,angle=270]{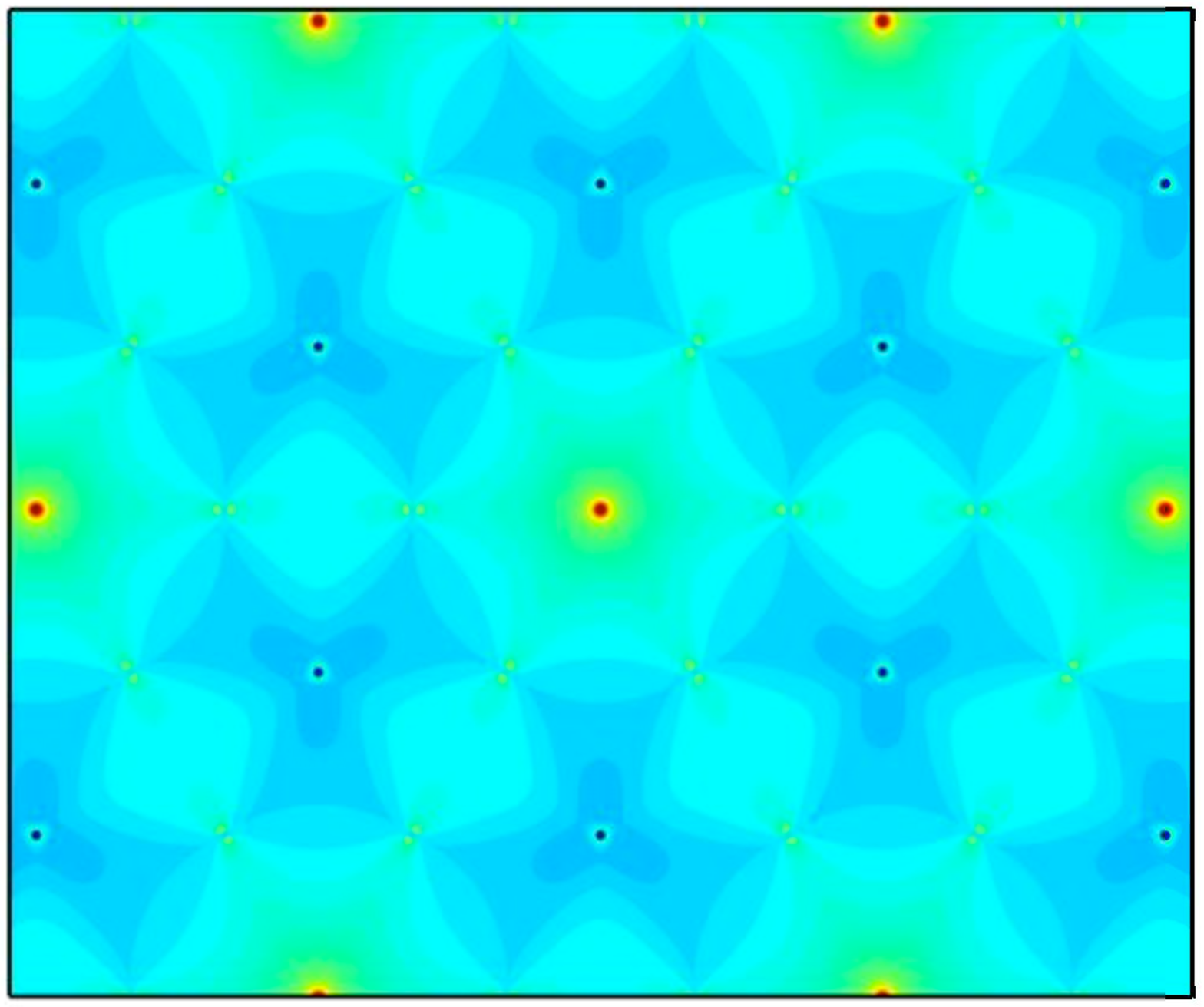}}
\hfil 
\subfigure[c)]{\includegraphics[width=0.28\textwidth,angle=270]{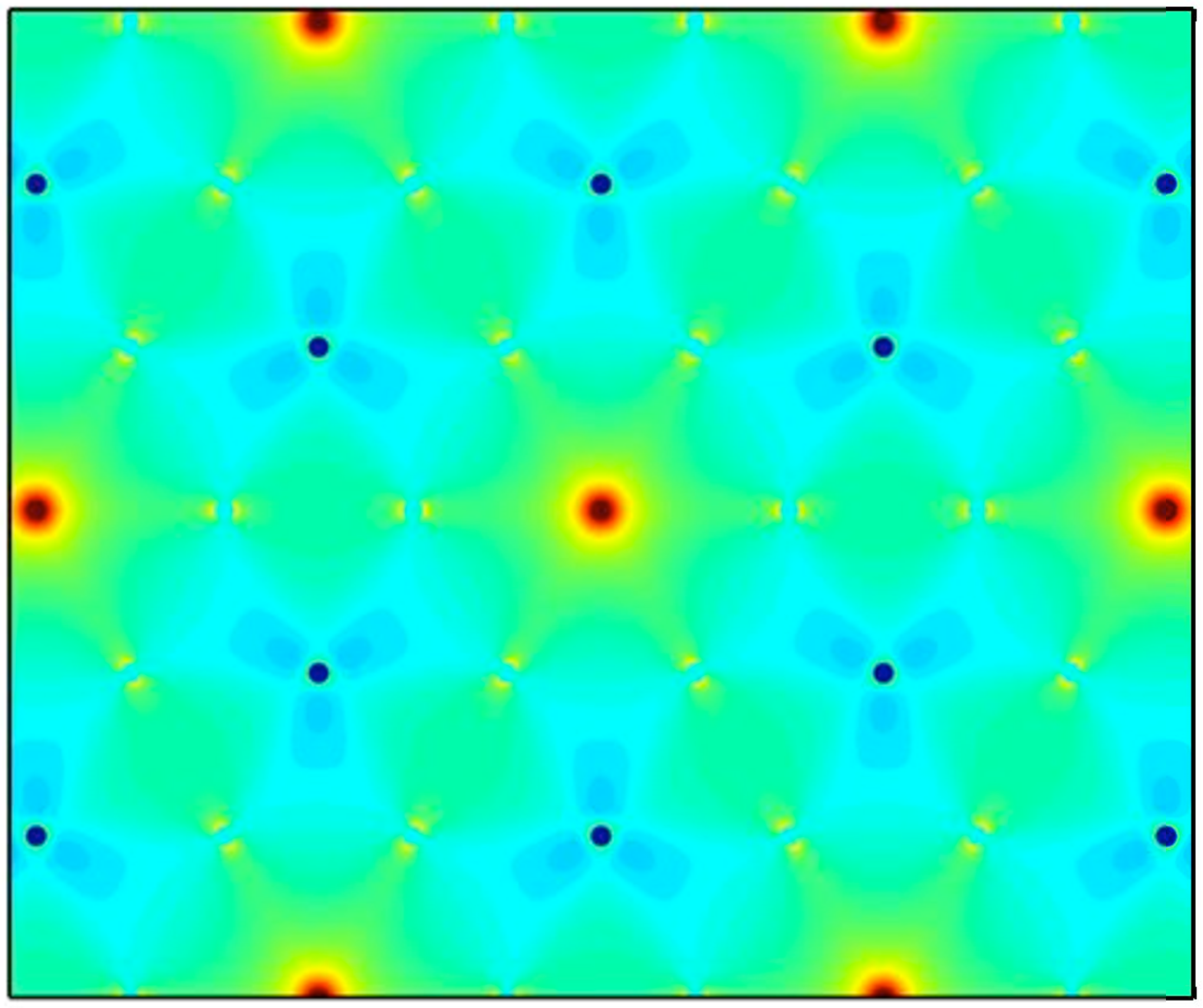}}
\hfil 
\subfigure[d)]{\includegraphics[width=0.28\textwidth,angle=270]{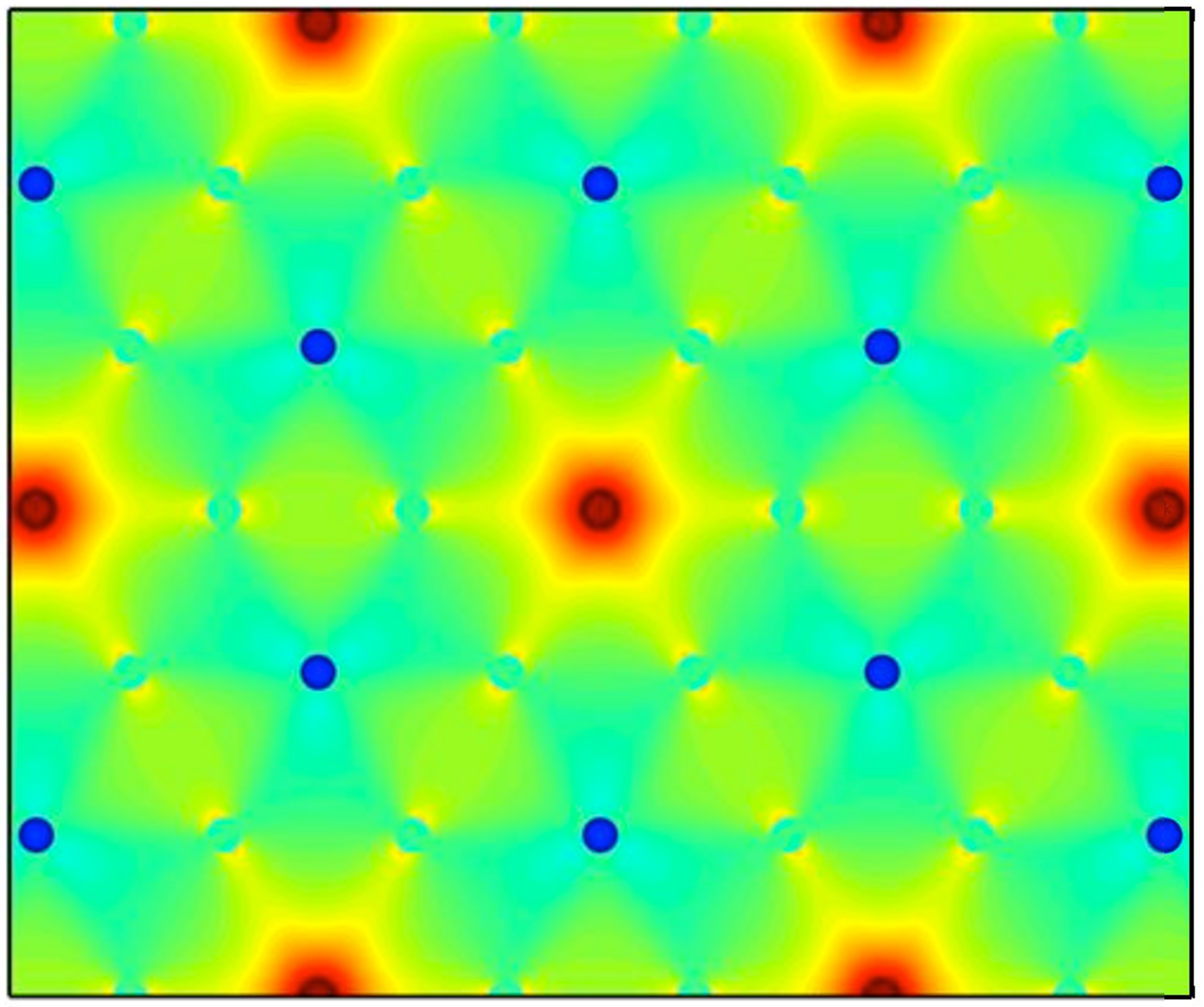}} 
\end{center}
\caption{Fourier transform of the local density of states for $t_0 = 2$ eV, $\epsilon_i = -1$ eV. The first row is the density of states in sublattice A ($\rho_a$), the second row is the density of states in sublattice B ($\rho_b$), and the third row is the sum of the two, $\rho_a+\rho_b$. Column (a) is for $\omega = 0.05$ eV. Column (b): $\omega = 0.15$ eV. Column (c): $\omega = 0.3$ eV. Column (d): $\omega = 0.5$ eV.}
\label{Fig:rho_kc}
\end{figure}
The column (a) of Fig. \ref{Fig:rho_k} refers to a vacancy. In this case, it is clear that
a $2q_F$ circumference is seen around the $\bm k=(0,0)$ point for the $\rho_a$ and $\rho_b$ plots, consistent with the modulations shown in Fig. \ref{Fig:ldos_R}. However, the intensity at the $2q_F$ circle around $\bm k = (0,0)$ is suppressed when looking at the $\rho_a + \rho_b$ plot. Features at six spots
at a distance $|K|$ around $\bm k=(0,0)$ are also seen, these correspond to the $K$ and $K^\prime$ points at the corners of the first Brillouin zone, and
represent inter-cone scattering. Additionally, six bright spots at distance $|K+K^\prime|$  are
also present. These vectors with modulus $|K+K^\prime|$ are reciprocal lattice vectors $\bm G$. 
For a pristine material
the local density of states has the periodicity of the underlying lattice, that is,
 $\rho(\bm r)=\rho(\bm r +\bm R)$,
and therefore, the Fourier transform of $\rho(\bm r)$ must show the same intensity
at $\bm k= (0, 0)$ and $\bm k=\bm G$. 
While the presence of an impurity breaks the periodicity of the real-space lattice, the
reason for $\bm k = (0,0)$ and $\bm k = \bm G$ being different is the fact that the impurity considered here is not on a whole unit cell, but on only one of the sites of the unit
cell. If there were only one site per unit cell and a
single short-range impurity, then the periodicity in
k-space would be maintained.  In the extreme case, if there was a line of impurities of the vacancy type, this would correspond to cutting the system into half, and $\bm k = (0,0)$  and $\bm k = \bm G$  would still be the same. Note that in all our cases $\rho_A$ is still periodic in $k$-space. The impurity on a site $A$ that we consider here, introduces structure within the $\bm R = 0$ unit cell, therefore, in k-space there is information going beyond the first Brillouin zone.
Mathematically this appears because the B-site is located at
$\bm R+ \bm \delta_3$, where $\bm R$ denotes the position of the $A$ sites, and $\bm \delta_3$ is not a lattice vector of triangular Bravais lattice (see Fig. \ref{Fig_lattice}). When we take the Fourier transform, the periodicity in $k$-space does
not happen at $\bm k = \bm G$ anymore, but at at larger $\bm k$.
All figures
show this signature, which is rather clear in column (c) of Fig. \ref{Fig:rho_k},  since
the Fermi surface energy has been chosen as large as $\omega=-1.5 $ eV.

Inter-cone scatterings, represented by the region around the $K$ and $K^\prime$ points, are highly angular-dependent \cite{bena2,barnea}, as can be seen particularly in the $\rho_b$ plots. While the scattering around the $\bm G$ vectors do not have such a strong angular dependence, some trigonal warping is observed in $\rho_B$. The scatterings around $\bm k = (0,0)$ are rotationally symmetric. Fig. \ref{Fig:rho_kc} corresponds to the case $t_0 = 2$ eV and $\epsilon_i = -1$ eV (case (c) of Fig. \ref{Fig:rho_k}), showing how the $k$-space LDOS map evolves with increasing energy. 

\section{STM current}
\label{secSTM}

In this Section we present calculations of the tunneling current between the STM tip and graphene, when the tip is close to an impurity atom.
We model the tip by the multimode tight-binding model, as described in Sec. \ref{tip}.
This choice departs from the more simplified approach where the
tip is modeled by a one dimensional system\cite{Mujica,Hong}.

There is a number of ways one can use to describe the tunneling of the electrons between the STM tip and graphene. Here we assume that
the coupling is made directly either to the impurity atom or to
the next neighbor carbon atom. This choice corresponds to probing the local
electronic properties at or around the impurity.
More general types of coupling are easily included
in the formalism. We write this coupling as
\begin{equation}
H_T=-W_2[c^\dag(0)d(0)+d^\dag(0)c(0)]\,, 
\end{equation}
where the operator $d(0)$ can represent either an electron at the impurity atom
in the $A$ sub-lattice or at the carbon atom in the $B$ sub-lattice.

Since the Hamiltonian of the problem is bilinear we can write it
in matrix form as 
\begin{equation}
 H=
\left[
\begin{array}{ccc}
H_b & V_L & 0\\
V^\dag_L & H_0 & V^\dag_R\\
0 & V_R & H_g
\end{array}
\right]
\label{hamiltonian33}
\end{equation}
where the matrices $V_L$ and $V_R$  represent the coupling of the last atom in the tip of
the STM microscope to the bulk of the tip and to graphene, respectively,
and $H_b$ and $H_g$ stand for the bulk Hamiltonians of the tip and
of graphene, respectively. $H_g$ also includes the impurity terms Eqs. (\ref{Vi}) and (\ref{V0}).  
The matrix $H$ is of infinite dimension due to $H_b$ and $H_g$.
The matrices $V_L^\dag$ and $V_R^\dag$ have the explicit form
\begin{equation}
 V_L^\dag=[\bm 0,-W_1,-W_1]\,,\hspace{0.5cm}V_R^\dag=[-W_2,\bm 0]\,,
\end{equation}
 where $\bm 0$ represents an infinite dimensional null row vector.

The tunneling is a local property, controlled by the coupling of the last atom of the tip
to the bulk atoms and to graphene. 
Since we want to compute local quantities, this
is best accomplished using Green's functions in real space. The full Green's function of the system
is defined by
\begin{equation}
(\bm 1 E+i0^+-H)G^+=\bm 1\,, 
\end{equation}
where  $\bm 1$ is the identity matrix. The matrix form of the Green's function is  
\begin{equation}
G^+= \left[
\begin{array}{ccc}
 G_{bb}  & G_{b0} & G_{bg}\\ 
G_{0b}  & G_{00} & G_{0g}\\
G_{gb}  & G_{g0} & G_{gg}
\end{array}
\right]\,.
\end{equation}
The quantity of interest is $G_{00}$, which can be shown to have the form 
\begin{equation}
G_{00}^+= (E+i0^+-\epsilon_0-\Sigma_L^+-\Sigma_R^+)\,, 
\end{equation}
where the matrices $\Sigma_L^+$ and $\Sigma_R^+$ are the self energies and have the form
\begin{equation}
 \Sigma_L^+=2W_1^2(G_{diag}+G_{offd})\,,
\hspace{0.5cm}
\Sigma_R^+=W_2^2G^+_{xx}\,,
\end{equation}
where $G^+_{xx}$ is the surface Green's function
of the Hamiltonian $H_g$ at the impurity unit cell
($x=a,b$), respectively.  Note that the quantity $G^+_{xx}$
is computed using Eq. (\ref{realspaceG}) and setting $\bm r=0$.

\begin{figure}[ht]
\begin{center}
\includegraphics*[angle=0,width=8cm]{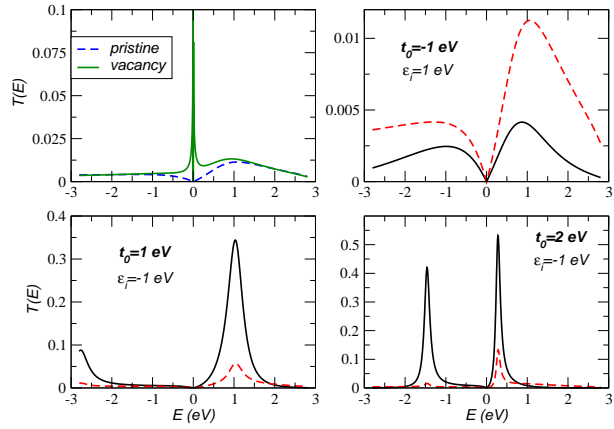}
\caption{Transmission probability $T(E)$.
The parameters used are (all in electron-volt):
$V=2$, $V_\perp=1$, $W_1$=0.9, $W_2=0.2$, and $\epsilon_0=0.2$.
The resonances seen in the LDOS in Fig. \ref{Fig:ldos_adatom}
show up in the transmission function.
The values of $t_0$ and $\epsilon_i$ are the same used in
Fig. \ref{Fig:ldos_adatom}
and the four panels here correspond to the same ones in that
figure.
 \label{Fig:TE_zero_V}}
\end{center}
\end{figure}

The study of non-equilibrium transport is done using the non-equilibrium 
Green's function method, or Keldysh formalism. This method is particularly suited to study the regime where the system has a strong departure from equilibrium, such as when the bias potential
on the STM tip, $V_b$, is large. In this work, we consider, however, that the system is in the steady state.  Since the seminal paper of Caroli
{\it et al.} on non-equilibrium quantum transport \cite{Caroli}, that the
method of non-equilibrium Green's functions started to be generalized to the
calculation of transport quantities of nanostructures.
There are many places where one can find
a description of the method \cite{Ferry,Jauho}, 
but a recent and elegant one was introduced
in the context of transport through systems having bound states, showing that
the problem can be reduced to the solution of an equation similar to a quantum Langevin equation\cite{Sen}. The general idea of this method is that two perfect leads are coupled to the system,
which is usually called the device. In our case the device is defined by the last
atom of the tip of the microscope. The Green's function of the
device has to be computed in the presence of the bulk of the tip and of graphene. 
This corresponds to our $G_{00}^+$ Green's
function. Besides the Green's function we need the effective coupling between the last atom
of the tip and the bulk atoms as well as that to the graphene atoms, 
which are determined in terms of the self-energies
\begin{equation}
\Gamma_{L/R} = \frac {i}{2\pi}(\Sigma^+_{L/R}-\Sigma^-_{L/R})\,. 
\end{equation}
Therefore the effective coupling $\Gamma_{L/R}$ depends on the surface Green's
function of the tip and of graphene. According to the general theory, the two systems
(bulk of the tip and graphene) are in thermal
equilibrium at temperatures $T_{L/R}$ and chemical potential $\mu_{L/R}$ and are connected
to the system at some time $t_0$. 
The total current through the
device is then given by 
\begin{equation}
J=\frac {2e}{h}\int_{-\infty}^\infty dE T(E)[f(E,\mu_L,T_L)-f(E,\mu_R,T_R)]\,, 
\end{equation}
where the factor of 2 is due to the spin degrees of freedom, $f(x)$ is the Fermi-Dirac distribution and
the transmission $T(E)$ is given by 
\begin{equation}
 T(E)= 4\pi^2\Gamma_L\vert G^+_{00}\vert^2\Gamma_R\,.
\end{equation}
In Fig. \ref{Fig:TE_zero_V} we depict $T(E)$ in different cases.
Note the asymmetry of the density of states which is exhibited even by the pristine case (Fig. \ref{Fig:TE_zero_V}, upper left panel). This asymmetry has a two fold nature: (i) it comes from the
fact the bulk of the tip has two transverse atoms but the tip has only one; (ii)
the fact that the atom at the tip has a different local energy from those in the
bulk. This asymmetry carries on to the disordered cases. Additionaly, for the
disordered cases the resonances seen in the local density of states has a strong
impact on the transition probability $T(E)$, leading to open transport
channels with large values of $T(E)$. This is specially true for the vacancy
and for the weakly coupled impurity case, that is, when hopping between impurity and carbon atoms is suppressed in relation to the hopping between carbon atoms, as expected for nitrogen substitution.
\begin{figure}[ht]
\begin{center}
\includegraphics*[angle=0,width=8cm]{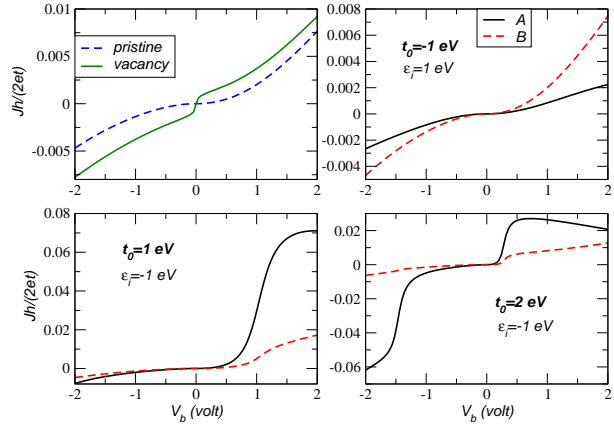}
\caption{STM current $J$.
The parameters used are (all in electron-volt):
$V=2$, $V_\perp=1$, $W_1$=0.9, $W_2=0.2$, and $\epsilon_0=0.2$.
The values of $t_0$ and $\epsilon_i$ are the same used in
Fig. \ref{Fig:ldos_adatom}.
The transmission function was computed at finite bias; the zero bias
case is given in Fig. \ref{Fig:TE_zero_V}.
The four panels here correspond to the same ones in that
figure \ref{Fig:TE_zero_V}.
 \label{Fig:J_zero_mu}}
\end{center}
\end{figure}

Since we want to probe the properties of the STM current at
zero doping we choose $\mu_L=eV/2$ and $\mu_R=-eV/2$. Also
$T_L=T_R=0$. This renders the calculation of the current to a simple
one-dimensional integral of $T(E)$ over the energy. The form
of the current will reflect the properties of $T(E)$ as function
of energy, and, as we have seen, these are markedly different
for the different cases, depending strongly on the value and sign
of $t_0$. The presence of resonances in $T(E)$ leads to steps
in the STM current. This is seen in Fig. \ref{Fig:J_zero_mu} for
the cases of the vacancy and to the case of weak coupling (positive $t_0$)
between
the impurity and the neighboring carbon atoms.

To fully characterize the STM current, another important quantity 
is the shot noise\cite{Beenakker}. For interacting systems, this
quantity contains information on the nature of the quasi-particles, including for example the possible existence of quasi-particles
with fractional charge. In disordered systems with no interactions,
information on transport through open channels can be obtained. For 
non-interacting electrons at zero temperature the
shot noise is defined as\cite{Buttiker}
\begin{equation}
 S=\frac{2e^2}{\hbar}\int_{\mu_R}^{\mu_L}dE\;T(E)[1-T(E)]\,.
\end{equation}
The relevant quantity is not $S$ directly but the Fano factor\cite{Beenakker}
defined as
 $F=\frac {S}{eJ}$.
When the transmission $T(E)$ is strongly reduced we have $F\rightarrow 1$, and 
the noise is said to be Poissonian. On the other hand, if the system has a finite
density of open channels, $T(E)\rightarrow 1$, we have $F<1$ due to $[1-T(E)]\ll 1$. 
The resonances that we obtain in $T(E)$ play a role in the resulting form of $F$. They lead to
an enhancement of the current, and to a significant decrease of the Fano factor
due to the opening of a transport channel. Note that the opening of the
channels is a consequence of the local disorder induced by the impurity.

\begin{figure}[ht]
\begin{center}
\includegraphics*[angle=0,width=8cm]{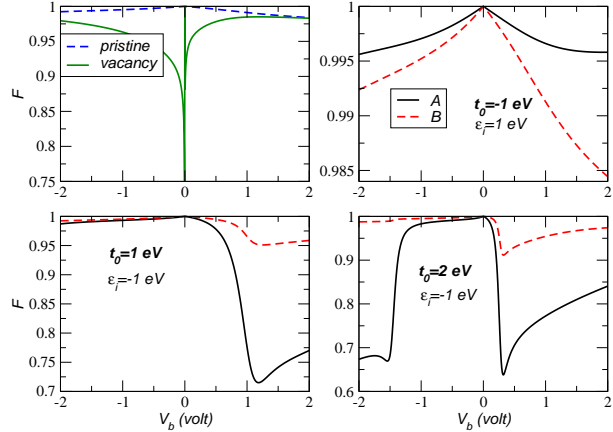}
\caption{Fano factor $F$.
The parameters used are (all in electron-volt):
$V=2$, $V_\perp=1$, $W_1$=0.9, $W_2=0.2$, and $\epsilon_0=0.2$.
The values of $t_0$ and $\epsilon_i$ are the same used in
Fig. \ref{Fig:ldos_adatom}. 
The four panels here correspond to the four panels of the current
given in Fig. \ref{Fig:J_zero_mu}.
 \label{Fig:Fano_zero_mu}}
\end{center}
\end{figure}
\section{Conclusions}
\label{secCONC}

In this paper we have studied the STM currents through locally disordered
graphene. We have considered a tip with transverse modes. Although the
tip is strictly quasi-one-dimensional it still departs from the
widely used model of a strictly one-dimensional model. Generalizing now the calculations
to a truly three-dimensional tip is reasonably straightforward. The modifications
would require introducing a three dimensional square lattice for describing the
bulk of the tip, and a decreasing number of atoms for each transverse
plane to describe the part of the tip in contact with graphene. 
This last part would lead to the most
significant change in the calculation, since the device would not be a single
atom as in our calculations here, but would be represented by a finite number
of them, and therefore the Green's function for the device would be a matrix
instead of a $c$-number. Nevertheless, as long as we take the dispersion
of the electrons in the tip to have large bandwidth, the current should not depend
much on the local density of states of the tip, since this would be essentially
constant. This corresponds to the usual wide band limit.

We have also seen that tunneling through either impurity atoms, or
their neighboring carbon atoms, depends on the local density of states
of graphene. For certain circunstances -- vacancy or weakly coupled impurities --
there is a development of resonances at or close to the Dirac point. These
resonances lead to a strong enhancement of the tunnelling probability which
appear as steps in the tunneling current. It is conceivable that graphene
could be locally modified in order to take advantage of these strong
resonances developed close to the Dirac point. Clearly the substituting
atoms would also locally distort the graphene lattice, an effect not included
in our description. How much the STM current would depart from the
values computed here would depend on the change in the values of the hopping
parameter, and on additional features on the density of states due
to the disorder. If future research will pursue the route of modifying
graphene locally, our results will be important for the characterization of the
surface. Even in the present state of affairs, our results could be used
to interpret STM current due to local impurities.

\ack
The authors acknowledge helpful discussions with A. H. Castro Neto
and J. M. B. Lopes dos Santos, and use of code shared by Filippos D. Klironomos.
NMRP acknowledges financial support from POCI 2010 via project
PTDC/FIS/64404/2006. 
SWT acknowledges partial support from UC-Lab FRP under award number 09LR05118602.

\section*{References}


\end{document}